\documentclass{aa}

\usepackage{graphicx}
\usepackage{txfonts}
\usepackage{amsmath}
\usepackage{amssymb}
\usepackage{wasysym}
\usepackage{color}
\usepackage{xcolor}
\usepackage{epsfig}
\usepackage{longtable}
\usepackage{pdflscape}
\usepackage{mathtools}
\usepackage{float}
\usepackage{footnote}
\usepackage{epstopdf}
\usepackage{units}
\usepackage{enumerate}
\usepackage{pdflscape}

\usepackage{txfonts}
\usepackage{subfiles} 

\newcommand{\kms}{km~s$^{-1}$}
\newcommand{\gaia}{\emph{Gaia}}
\newcommand{\rgc}{R_{\mathrm{GC}}}

\begin{document}

   \title{3D kinematics and age distribution of the Open Cluster population \thanks{The tables with star and cluster velocities are only available in electronic form at the CDS via anonymous ftp to cdsarc.u-strasbg.fr (130.79.128.5) 
or via http://cdsarc.u-strasbg.fr/viz-bin/qcat?J/A+A/?/?}}

   \author{Y. Tarricq\inst{1},
          C. Soubiran\inst{1},
          L. Casamiquela\inst{1},
          T. Cantat-Gaudin\inst{2},
          L. Chemin\inst{3},
          F. Anders\inst{2}, 
          T. Antoja\inst{2},
          M. Romero-Gómez\inst{2}, 
          F. Figueras\inst{2},
          C. Jordi\inst{2}, 
          A. Bragaglia\inst{4}, 
          L. Balaguer-Núñez\inst{2}, 
          R. Carrera\inst{5},
          A. Castro-Ginard\inst{2}, 
          A. Moitinho\inst{6},
          P. Ramos\inst{2},
          D. Bossini\inst{7}}
    \authorrunning{Y. Tarricq et al.} 

   \institute{Laboratoire  d’Astrophysique  de  Bordeaux,  Univ.  Bordeaux,  CNRS,  B18N,  allée  Geoffroy  Saint-Hilaire,  33615  Pessac,  France\\
              \email{yoann.tarricq@u-bordeaux.fr}
         \and   
             Institut de Ciències del Cosmos, Universitat de Barcelona (IEEC-UB), Martí i Franquès 1, E-08028 Barcelona, Spain
         \and
             Centro de Astronomía, Universidad de Antofagasta, Avda. U.de Antofagasta 02800, Antofagasta, Chile 
         \and
             INAF-Osservatorio di Astrofisica e Scienza dello Spazio, via P. Gobetti 93/3, 40129 Bologna, Italy
         \and
	     INAF-Osservatorio Astronomico di Padova, vicolo dell’Osservatorio 5, 35122 Padova, Italy
         \and
             CENTRA, Faculdade de Ciencias, Universidade de Lisboa, Ed. C8, Campo Grande, 1749-016 Lisboa, Portugal
         \and
             CAUP - Centro de Astrofisica da Universidade do Porto, Rua das Estrelas, Porto, Portugal. 
             }

\date{\today}

 
  \abstract
   {Open Clusters (OCs) can trace with a great accuracy the evolution of the Galactic disk. \emph{Gaia} and large ground based spectroscopic surveys make it possible to determine their properties and to study their kinematics with an unprecedented precision.}
   {The aim of this work is to study the kinematical behavior of the OC population over time. We take advantage of the latest age determinations of OCs to investigate the correlations of the 6D phase space coordinates and orbital properties with age. The phase space distribution, age-velocity relation and action distribution are compared to those of field stars. We also investigate the rotation curve of the Milky Way traced by OCs and we compare it to that of other observational or theoretical studies.}
   {We gathered nearly 30\,000 Radial Velocity (RV) measurements of OC members from both \emph{Gaia}-RVS data and ground based surveys and catalogues. We computed the weighted mean RV, Galactic velocities and orbital parameters of 1\,382 OCs. We investigated their distributions as a function of age, and by comparison to field stars.}
   {We provide the largest RV catalogue available for OCs, half of it based on at least 3 members. Compared to field stars, we note that OCs are not exactly on the same arches in the radial-azimuthal velocity plane, while they seem to follow the same diagonal ridges in the Galactic radial distribution of azimuthal velocities. Velocity ellipsoids in different age bins all show a clear anisotropy. The heating rate of the OC population is similar to that of field stars for the radial and azimuthal components but significantly lower for the vertical component. The rotation curve drawn by our sample of clusters shows several dips, which match the wiggles derived from non-axisymmetric models of the Galaxy. From the computation of orbits, we obtain a clear dependence of the maximum height and eccentricity with age. Finally, the orbital characteristics of the sample of clusters as shown by the action variables, follow the distribution of field stars. The additional age information of the clusters points towards some (weak) age dependence of the known moving groups.}
   {}


   \maketitle

\section{Introduction}
The study of kinematical properties of the Milky Way open clusters (OCs) has a long tradition. Their motion can be used to understand the Milky Way's gravitational potential and the various perturbations which act on the structure and dynamics of the Galactic disc. The solar neighbourhood is known to be clumpy \citep[e.g. ][]{egg96,deh98,ant12}, and the relation of observed substructures with some OCs, in particular the Hyades, has been established \citep{egg58,che99}. While stellar streams were initially believed to be remnants of star clusters, their origin is now believed to be the result of the disruption of star clusters or of dynamical origin such as the resonant trapping by the bar and the spiral arms \citep{fam08} or the passage of the Sagittarius dwarf galaxy within the galactic plane \citep{mon18, kha19}. Stellar streams could also originate from a combination of the aforementioned processes that happen together. \emph{Gaia} DR2 data \citep{dr2} revealed the complexity of the local velocity distribution with an unprecedented resolution, showing in particular that the velocity distribution of nearby OCs overlaps well with prominent arched over-densities of moving groups \citep{kat18}.\\

\emph{Gaia} DR2 has considerably fostered the study of OCs with the determination of new memberships for an unprecedented number of stars and clusters. A first large catalogue was published by \cite{can18} who systematically looked for members around the $\sim$3300 catalogued OCs (mostly from \cite{dia02} and \cite{kha13} studies), only based on \emph{Gaia} DR2 positions, parallaxes and proper motions. Membership probabilities were computed for $\sim$400\,000 stars and among other parameters, the most probable distances were determined for 1229 OCs. A large fraction of candidate OCs could not be confirmed by \emph{Gaia} or was proved to correspond to chance alignment. The number of OC candidates has also significantly increased, either thanks to serendipitous discoveries in \emph{Gaia} DR2 \citep{can18,fer19}, or as the result of systematic searches \citep{cas18,cas19,cas20,can19,sim19,liu19}. An updated catalogue of membership probabilities  for 1481 OCs has been provided by \cite{can20a}. The physical and kinematical properties of OCs has also been revisited thanks to \gaia. Using the memberships from \cite{can18} and only \emph{Gaia} data, \cite{bos19} determined the age, distance modulus and extinction of 269 OCs, and \cite{sou18} computed the 6D phase-space information of 861 clusters.  \cite{car19} increased by 145 the number of OCs with full 6D phase-space information by looking for members in the GALAH and APOGEE spectroscopic surveys. The vertical distribution of young clusters was found to be very flat, with a dispersion of vertical velocities of 5 km s$^{-1}$, while clusters older than 1 Gyr span distances to the Galactic plane of up to 1 kpc with a vertical velocity dispersion of 14 km s$^{-1}$, typical of the thin disc. Recently, \cite{can20b} derived physical properties of all known OCs identified in the \emph{Gaia} data in a homogeneous fashion, with a method based on isochrone fitting and an artificial neural network, ending up with 1867 clusters with reliable ages. 

The paper is organized as follows. Sect. \ref{s:RV} describes the collection of individual RVs gathered from \emph{Gaia} DR2 \citep{sar18,kat19} and from ground-based resources, used to build two catalogues of mean RVs, per star and per cluster.
In Sect. \ref{sec:GalacticVelocities}, 6D galactocentric coordinates are computed. We compare the resulting distributions with those of nearby field stars. We evaluate how the different coordinates vary with the age of the clusters and we determine the age velocity relation of the population.  We investigate how OCs follow the theoretical rotation curve of the Galactic disc.  Orbital parameters and actions are described in Sect. \ref{action_angles_potentials} and used to revisit the link between OCs of different ages and the phase-space substructures of the disc. The conclusions of this study are summarized in Sect. \ref{sec:conclusion}

\section{Radial velocities}\label{s:RV}

\subsection{Input data}
We took advantage of the catalogue of OCs by \citet{can20b} which provides a list of probable members for 2017 OCs that they used to estimate ages. Their memberships mostly come from \citet{can20a} who used the unsupervised classification scheme UPMASK \citep{kro14,can18}. They also applied  UPMASK to the recently discovered University Barcelona Clusters \citep[UBC, ][]{cas18,cas19,cas20} and to those discovered by \cite{liu19}. For the Hyades and Coma Berenices, they adopted the list of members published by \cite{bab18}, UPMASK being unable to recover members for populated clusters that are too extended on the sky. 

Based on the aforementioned list of members, containing $\sim$475\,000 stars, we gathered all the RV measurements available for them in various surveys and catalogues. Our main source was \emph{Gaia} DR2 which includes RVs for about 7 million stars \citep{sar18,kat19}.
We also queried several catalogues from large spectroscopic surveys in addition to \emph{Gaia}: the latest public version of \emph{Gaia}-ESO survey \citep{ges}, APOGEE DR16 \citep{apo}, RAVE DR6 \citep{rave}, and GALAH DR3 \citep{galah}. We did not consider LAMOST \citep{cui12, lamost2} because its RV precision and accuracy ($\sim$5 \kms) are not at the same level.  We included radial velocities derived in by the OCCASO survey \citep[][Carrera et al. in preparation]{casa2016}. We also considered the RV catalogues by \cite{cs18,mer09,mer08, wor12, nor04}. Several quality cuts were applied to individual measurements. For RAVE we applied the criteria suggested by \cite{rave} by selecting stars verifying |correctionRV| $<$10 \kms, $\sigma_{\rm RV} < 8$ \kms\ and correlationCoeff $>$10. For APOGEE, we rejected the stars flagged with VERY\_BRIGHT\_NEIGHBOR, BAD\_PIXELS or LOW\_SNR as recommended in the online documentation of the survey. For \gaia\ RVS we filtered the erroneous RVs found by \cite{bou19}. Despite these cuts, there were still individual measurements with large uncertainties incompatible with the required precision for this work. This convinced us to filter out all the individual RVs having uncertainties larger than 8 \kms, the same cut as for RAVE. The rejected values represent 6\% of the full set, 10\% for the RVS set. Then we noticed 43 stars with $|{\rm RV}| > 200$ \kms\ which turned out to be mainly OB-type or Wolf-Rayet or spectroscopic binary stars according to Simbad. Such stars with unreliable or variable RVs were rejected.

Table \ref{t:rv_cat}gives the number of cluster members retrieved in each of the catalogues after filtering, with the median uncertainty of the corresponding RVs, as quoted in the catalogues. In \cite{cs18}, the uncertainty corresponds to the standard error of the weighted mean for stable stars that have been followed-up for exoplanet detection. The high precision of individual measurements and number of observations explain the very low median uncertainty of that catalogue (0.002 \kms). APOGEE, OCCASO and the catalogues from \citet{mer09,mer08} and \citet{nor04} also contain stars that have been observed several times in order to identify binaries. In that case, the RV uncertainty corresponds to the quadratic sum of the single measurement error with the scatter of the measurements. RAVE also has multiple observations for a small fraction of stars and provides the individual values with their uncertainty corresponding to the error of a single measurement. Finally the \emph{Gaia}-ESO survey includes RVs of the same star obtained with different setups that give different measurement errors \citep{jac15}. 

\begin{figure*}
\centering
\includegraphics[width=\textwidth]{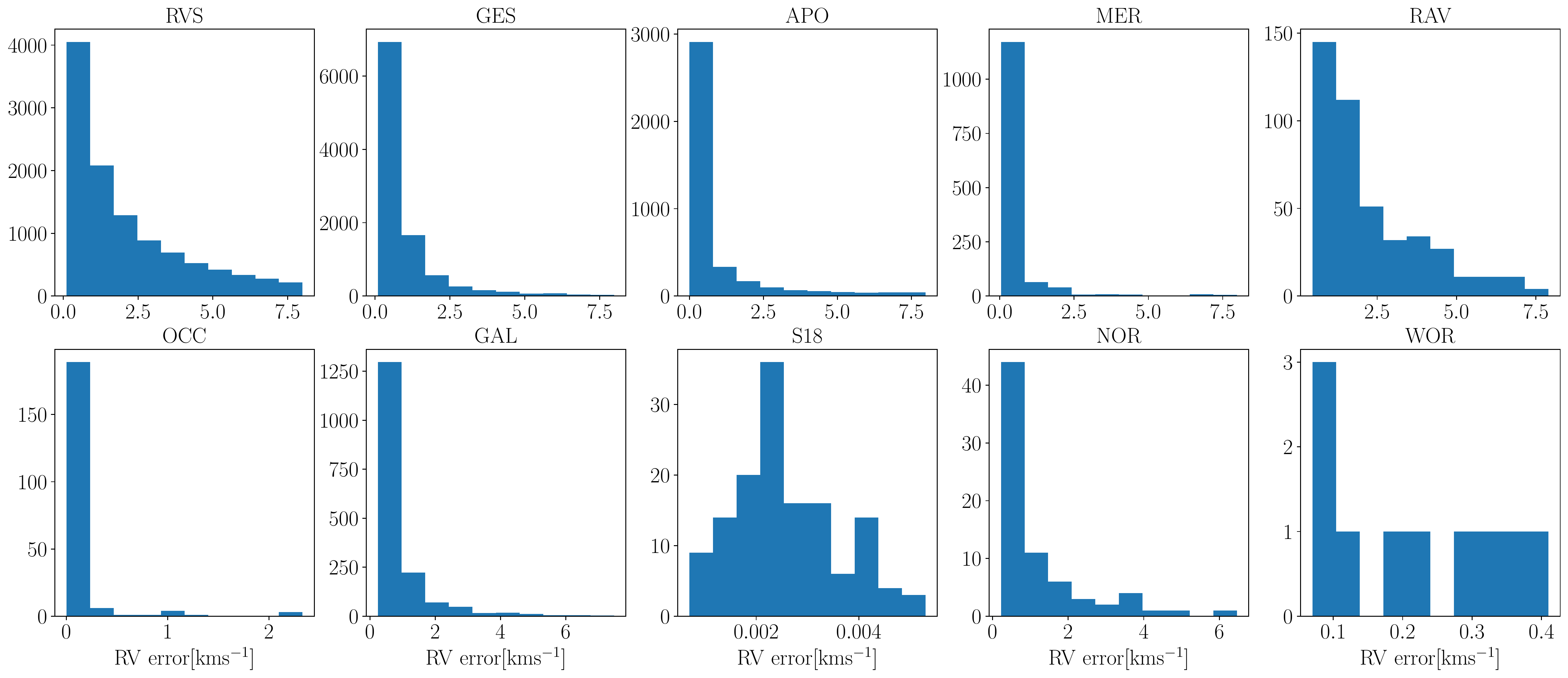}
\caption{Distribution of the radial velocity uncertainties of each catalog considered in this study and designated by short names as listed in Table  \ref{t:rv_cat}}
\label{fig:error_distrib}
\end{figure*}

This star sample is dominated by dwarfs and has a median effective temperature Teff of 4874 K (based on Teff provided in \gaia\ DR2), with more than 13\% of the stars being hotter than 6250 K. We note that uncertainties of stars hotter than 6250 K are significantly larger (median 1.75 \kms) than that of cooler stars (median 0.53 \kms). This mainly reflects the lower precision of the RVs of rotating stars, which are frequent among AF-type stars in young OCs. Fig. \ref{fig:error_distrib} shows the distribution of the RV uncertainties of each catalogue. Note that uncertainties are not determined the same way in the different catalogues. The  28\,371 RV measurements of OCs members that we gathered correspond to 23\,453 different stars.

\begin{table}[h]
  \centering 
  \caption{Number of stars in common between the catalogue of \cite{can20b} and catalogues of radial velocities, with the corresponding median uncertainty (in \kms), after the quality cuts described in the text. Catalogues are designated by short names as defined below the table.}
  \label{t:rv_cat}
\begin{tabular}{l | r | c}
\hline
  survey / catalogue    & $N_{\mathrm{stars}}$ & median RV error \\ 
\hline
RVS  & 10741 & 1.33 \\
GES  & 9894  & 0.50 \\
APO  & 3212  & 0.21 \\
GAL  & 1724  & 0.59 \\ 
MER  & 1313  & 0.28 \\
RAV  & 386   & 1.59 \\
OCC  & 187   & 0.04 \\
S18  & 138   & 0.002 \\
NOR  & 73    & 0.60 \\
WOR  & 10    & 0.20 \\
\hline
\end{tabular}\\
\flushleft
RVS: \emph{Gaia} DR2;
GES: \gaia\ ESO survey;
APO: APOGEE DR16;
GAL: GALAH DR3;
MER: \cite{mer08,mer09};
RAV: RAVE DR6;
OCC : OCCASO \cite{casa2016};
S18: \cite{cs18};
NOR: \cite{nor04};
WOR : \cite{wor12}
\end{table}

\subsection{Zero points of RV catalogues}

\begin{figure*}
\centering
\includegraphics[scale=0.5]{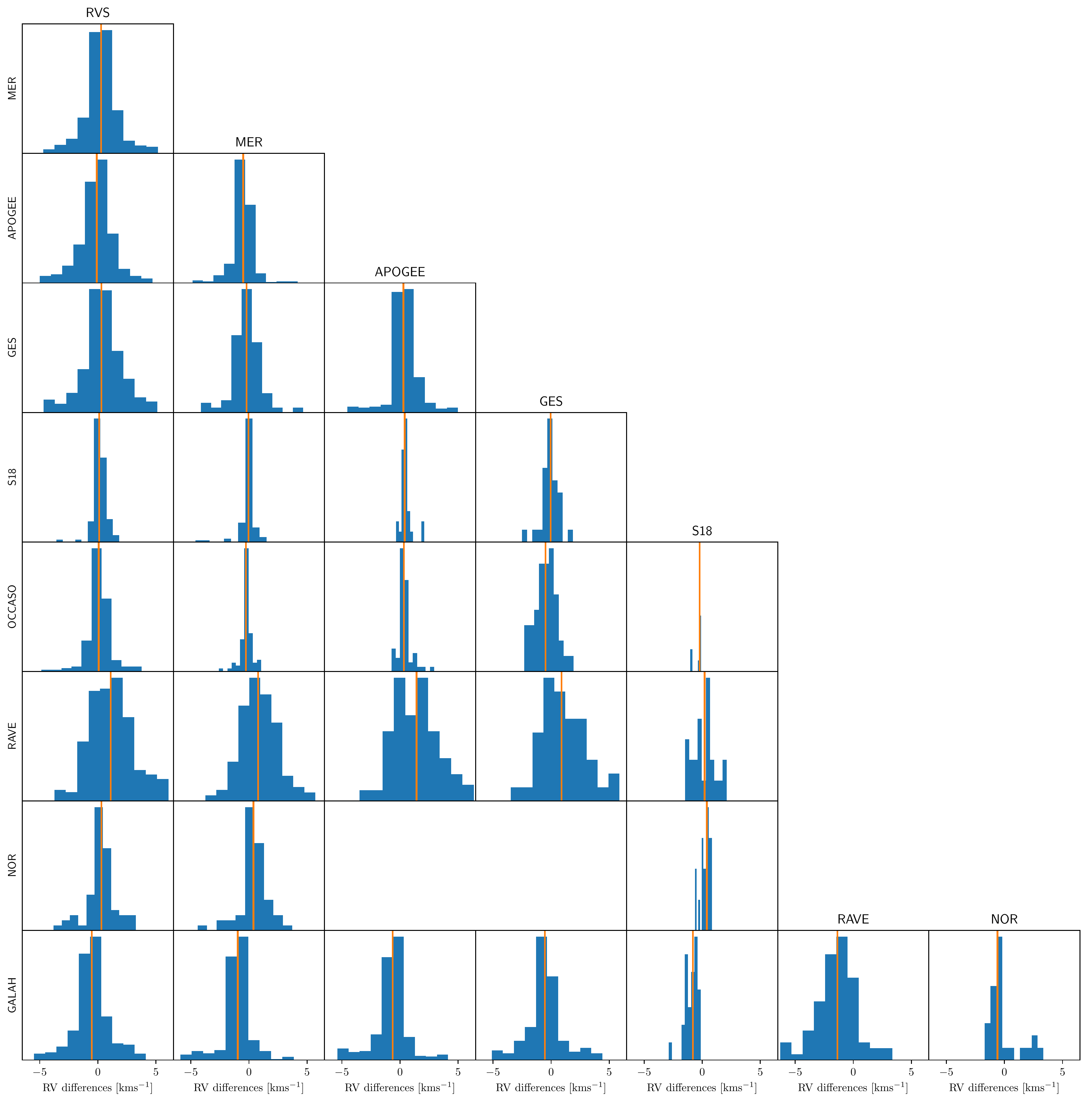}
\caption{Histograms of the RV difference of common stars in several surveys (zoomed in the range from -5 to 5 \kms). The orange solid line corresponds to the median value. The statistics of the comparisons are given in Table \ref{t:diff_stat}.}
\label{fig:residus}
\end{figure*}

This sample allowed us to assess the consistency of RVs in the different catalogues. The comparison of RVs for stars in common in two catalogues gives an idea of potential offsets due to zero-point differences, together with their typical precision. Zero-point differences are a result of the different observing modes, instrumental characteristics and calibration procedures of each instrument. They have to be taken into account when combining RVs of different origins. A subset of 3\,116 stars have measurements in two or more catalogues. The RV difference between catalogues that have more than 20 stars in common is presented in Fig. \ref{fig:residus}, and the corresponding comparison among catalogues in Table \ref{t:diff_stat}. The RVs have a general good agreement with offsets smaller than 0.5 \kms\, and between 0.5 and 1.4 \kms\ for comparisons that involve RAVE or GALAH. The dispersions (measured by the median absolute deviation MAD) are typically of the order 1 \kms  or below, consistent with the precision of the catalogues listed in Table \ref{t:rv_cat} for all combinations of surveys. 
In these comparisons of the different catalogues, we did not see any significant trend with colour or apparent magnitude.


\begin{table}[h]
  \centering 
  \caption{Median (MED) and MAD of RV differences for stars observed by two surveys with the number of stars in common. Short names of the catalogs are the same as in Table \ref{t:rv_cat}.}
  \label{t:diff_stat}
\begin{tabular}{lrrr }
\hline
  catalogues    & $N_{\mathrm{stars}}$ & MED & MAD \\ 
\hline
RVS - MER & 1079 & 0.28 & 0.68 \\
RVS - APO & 784 & -0.09 & 0.86 \\
RVS - GES & 486 & 0.31 & 1.06 \\
RVS - S18 & 115 & 0.12 & 0.26 \\
RVS - OCC & 161 & 0.07 & 0.40 \\
RVS - RAV & 299 & 1.12 & 1.36 \\
RVS - GAL & 607 & -0.51 & 0.88 \\
RVS - NOR & 65 & 0.31 & 0.49 \\
MER - APO & 304 & -0.50 & 0.33 \\
MER - GES & 152 & -0.21 & 0.72 \\
MER - S18 & 106 & -0.03 & 0.16 \\
MER - OCC & 82 & -0.26 & 0.17 \\
MER - RAV & 198 & 0.79 & 1.13 \\
MER - GAL & 293 & -0.96 & 0.49 \\
MER - NOR & 63 & 0.39 & 0.47 \\
APO - GES & 358 & 0.29 & 0.43 \\
APO - S18 & 30 & 0.40 & 0.10 \\
APO - OCC & 65 & 0.33 & 0.13 \\
APO - RAV & 125 & 1.42 & 1.60 \\
APO - GAL & 534 & -0.64 & 0.59 \\
GES - S18 & 30 & -0.04 & 0.41 \\
GES - OCC & 41 & -0.48 & 0.66 \\
GES - RAV & 57 & 0.89 & 1.89 \\
GES - GAL & 281 & -0.54 & 0.78 \\
S18 - OCC & 21 & -0.22 & 0.04 \\
S18 - RAV & 24 & 0.20 & 0.71 \\
S18 - GAL & 29 & -0.80 & 0.36 \\
S18 - NOR & 20 & 0.40 & 0.23 \\
RAV - GAL & 97 & -1.37 & 1.43 \\
GAL - NOR & 28 & 0.60 & 0.27 \\
\hline
\end{tabular}
\end{table}

Most of the RVs available for the cluster members are provided by \emph{Gaia} DR2 or the \emph{Gaia} ESO survey. GES has indeed a large fraction of its observing program dedicated to OCs and nicely increases the number of clusters for which a RV can be computed. \emph{Gaia} and GES have an offset of $\sim$0.3 \kms, similar to the offset between GES and APOGEE, while \emph{Gaia} and APOGEE agree at a level better than 0.1 \kms. 
The zero-point of RAVE is found to be different from the other surveys by $\sim$1 \kms\ (RAVE underestimates the RVs compared to the others) which is larger than what was reported in previous studies. \cite{ste18, rave} and \cite{sar18} report an offset of $\sim$0.3 \kms\ between RAVE and \gaia\ DR2. Here, even when we apply strong quality cuts on both RAVE and \gaia\ DR2, the median difference remains 1 \kms. RAVE also differs from APOGEE by 1.42 \kms, and from \cite{mer08,mer09} by 0.79 \kms. The agreement with S18 is, however, better at a level of 0.2 \kms, possibly related to the bright magnitude of the common stars and their colour range. We also note a systematic offset of GALAH DR3 with all the other catalogues, opposite to that of RAVE and of smaller amplitude, but still larger than the value of 0.22 \kms\ reported by \cite{galah} for the comparison to \gaia\ DR2. These larger offsets are possibly related to the fact that our star sample is dominated by dwarfs, with a fraction of early-type stars larger than in RAVE and GALAH in general. For RAVE, we see indeed a trend of the offset being slightly larger for hot stars than for cool stars.

In order to put all the RVs on the scale of \gaia\, we applied a zero-point correction to the individual RVs from all the non-\gaia\ catalogues, according to the offsets listed in Table \ref{t:diff_stat}.

\subsection{Mean radial velocities}

In order to compute the mean RV of each OC, we first computed the mean RV of each star, since some of them (12\%) have multiple measurements from different catalogues. For both the mean per star and the mean per cluster, we used a weighted procedure based on the errors of individual measurements, following \cite{sou13,sou18}. For each star or each cluster, the mean RV is computed by attributing to each RV measurement a weight $\omega_{i}$ defined as $\omega_{i} = 1/\epsilon_{i}^{2}$, $\epsilon_{i}$ being the RV error. Outliers were rejected on the basis of a 3$\sigma$ clipping. A fraction of about 8\% among stars with multiple measurements have a RV uncertainty larger than 3 \kms\ and we suspect that they are binaries with variable RV. Such stars are not rejected but the procedure gives them less weight when the mean RV of their parent cluster is computed. However, in the large majority of cases, we have only one measurement per star so that binaries cannot be identified. Binaries that have large variations of their RV may alter the mean RV of the parent OC in case of few members.

The catalogue of stars provides the mean RV in the \gaia\ RVS scale, with its uncertainty and the number of measurements, as well as the membership probability from \cite{can20b}. It includes 23\,424 unique stars, with 97 of them appearing twice owing to a non-null probability to belong to two different clusters being close to each other on the sky.



In order to compute the mean RV per OC, we considered only the stars with a membership probability higher than 0.4, the threshold value found by \cite{sou18} to be the best compromise between the largest number of members and the lowest contamination by field stars. 
In the end, 1\,382 OCs have a mean RV, which represents an improvement by 60\% in size compared to the previous RV catalogue of OCs, based on \gaia\ only \citep{sou18}. It also supersedes the catalogue by \cite{car19} based on \gaia, GALAH and APOGEE. Before \gaia\ DR2 the two largest compilations of cluster RVs were those of \cite{kha13} and \cite{dia02} with respectively 962 and 703 objects. 

\begin{figure}[t]
\centering
\includegraphics[width=\columnwidth]{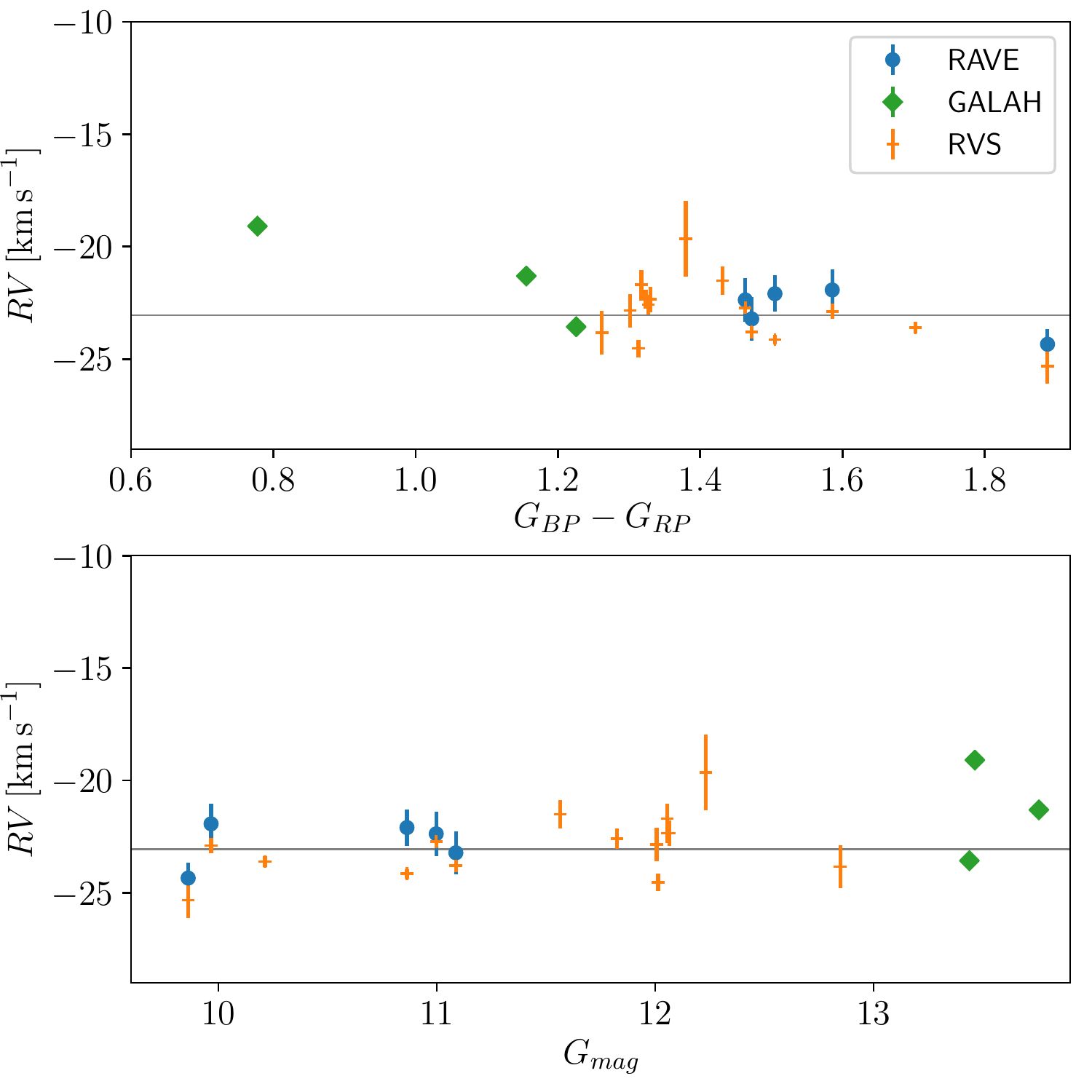}
\caption{Individual RV of UBC~274 members (proba $\ge$ 0.4), from RAVE in blue, from GALAH in green, both after the zero point correction indicated in Table \ref{t:diff_stat}, and from \gaia\ RVS in orange, as a function of the $G_{BP}-G_{RP}$ colour and $G$ magnitude from \gaia. The grey line represents the weighted mean, RV=-23.05$\pm$0.25 \kms, based on 18 members.}
\label{fig:UBC}
\end{figure}

This is the first determination of mean RV for most of the recently discovered OCs. In particular half of the UBC clusters \citep{cas20} are part of our catalogue, and they represent nearly 20\% of the full sample of OCs having a known RV. Among the 75 high confidence clusters recently discovered by \cite{liu19}, 35 are present in the latest catalog from \cite{can20b} and we provide a RV for most of them. For instance, we provide the mean RV of UBC~274 (also reported as LP~5), a recently discovered disrupting OC \citep{cas20,pia2020}, based on 18 members observed with RAVE, GALAH and \gaia\ RVS (Fig. \ref{fig:UBC}). Thanks to the combination of \gaia\ RVS and GES, several OCs have now their RV based on hundreds of members, the top two being Trumpler 5 and NGC 3532 with respectively 659 and 664 stars. For about 18\% of the sample, the mean RV is based on more than 10 stars, and for 50\% it is based on at least 3 stars. The median uncertainty of the weighted mean RV is 1.13 \kms\ when the full sample is considered.

Among the clusters with fewer members, some exhibit a large error which renders their mean RV uncertain. Some 430 OCs have a RV based on one star only, with no information on its potential variability (binarity). This represents 31\% of our sample, nearly the same proportion as in \citet{sou18}. Selecting the most reliable OCs which have a RV uncertainty lower than 3 \kms\ based on at least 3 stars, we get 513 clusters. This sub-sample has a median uncertainty of 0.55 \kms\ and a median number of stars of 9. This is 107 more reliable OCs than in \citet{sou18}.

There are 21 high velocity OCs in the catalogue with |RV| $>$ 100 \kms\ but 13 are based on a single star, therefore to be considered with caution. The 8 remaining high velocity clusters with more members were previously known.

The catalogue of OC velocities (released as online data) provides the RV per cluster with its uncertainty and the number of members used in the average, as well as galactic velocities, orbital parameters and actions computed in the next sections.


\section{Open Clusters in phase space}\label{sec:GalacticVelocities}
In this section, we combined the mean OC positions, distances, proper motions from \cite{can20b} and our mean RVs to compute heliocentric and Galactocentric cartesian and cylindrical positions and velocities for the full sample of 1382 OCs. For positions, the same conventions and reference values as \cite{kat18} were used: in cartesian Galactocentric coordinates, the Sun is located at $X=-8.34$ kpc, $Y=0$ pc and $Z=27$ pc from the centre of the Galaxy. Similarly, we set the azimuthal velocity $V_{c}$ at the solar radius at 240 \kms. The velocity of the Sun with respect to the local standard of rest (LSR) is set to $(U,V,W)=(11.1,12.24,7.25)$ \kms\ \citep{sch2010}. 


To calculate the uncertainties in the velocity space we used a Monte Carlo sampling of the astrometric and RV measurements and their uncertainties which we assumed Gaussian. In the following, we considered the standard deviations of the Monte Carlo samples as the uncertainties.
We then ended up for the full sample of OCs with median standard deviations in the cylindrical velocities of ($\delta V_{r},\delta V_{\phi},\delta V_{z}) = (2.8, 3.2, 1.6)$ \kms.

For distant OCs, the astrometric uncertainties translate into large Galactic velocity errors. There are 129 OCs with standard deviations on one of the velocity components higher than 100 \kms, all of them at distances from the Sun larger than 2.2 kpc. The most extreme cases are NGC~3105 and SAI~109 which have a relative parallax error of respectively 82 and 99\% and are both located at 6.9 kpc. For these two clusters, the velocity uncertainties reach several thousand \kms\ in $V_{\phi}$ and $V_{z}$. Also, as mentioned in the previous section, some OCs do not have a fully reliable mean RV because of few members, lack of information about their stability or a large dispersion among the members.  

We therefore defined a High Quality Sample (HQS) composed of the OCs with a reliable mean RV (uncertainty lower than 3 \kms\ based on at least three members as defined in the previous section) and having standard deviations on ($V_{r},V_{\phi},V_{z}$) below 10 \kms. This leaves us with a HQS composed of 418 OCs with median uncertainties on ($V_{r},V_{\phi},V_{z}$) of (1.2, 1.3, 1.0) \kms. This is very similar to the median uncertainty of the field star sample selected by \citet{kat18} who quoted ($\delta V_{r},\delta V_{\phi},\delta V_{z}) = (1.2, 1.3, 1.0)$ \kms\ for 
stars with $\varpi/\sigma_{\varpi}>5$. Out of these 418 clusters, an homogeneous estimation of their ages is provided by \cite{can20b} for 411 of them.

\begin{figure}[t]
\centering
\includegraphics[scale=0.4]{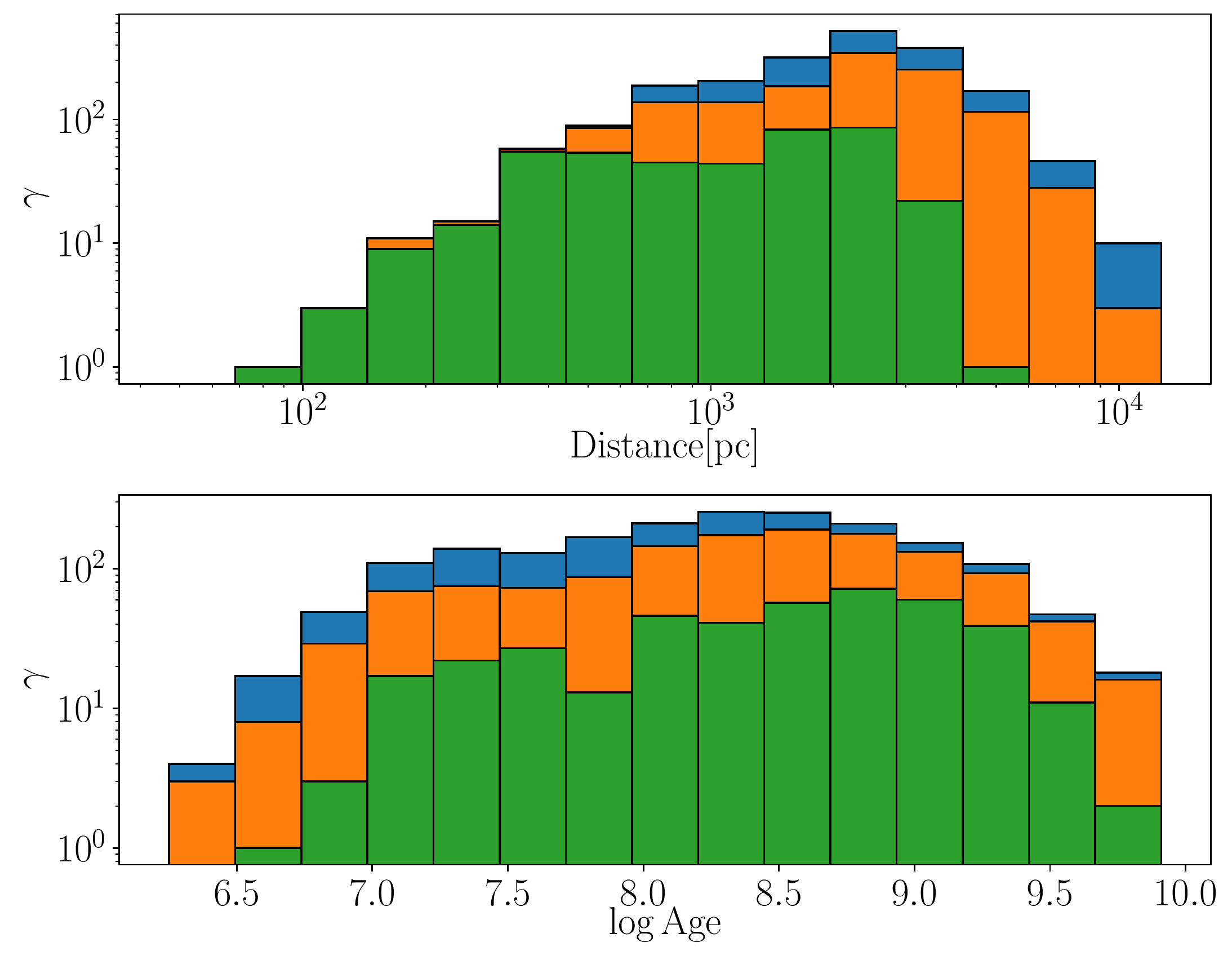}
\caption{Histogram in log scale of the heliocentric distance (top panel) and of the age distribution (bottom panel) for the full sample of OCs from \citet{can20b} in blue, the subsample of OCs for which we have a RV measurement in orange and the HQS in green.}
\label{fig:histo_samples}
\end{figure}

Figure \ref{fig:histo_samples} shows the histogram of the heliocentric distance and logarithmic ages of our different samples. Our subsample of 1382 OCs with RV measurements is 90\% complete with respect to the total sample of 2017 OCs from \citet{can20b} up to 860 pc while our HQS is 90 \% complete with respect with the total sample up to 500pc.

On the bottom panel, we can see that except for the very young clusters, all ages are represented in the HQS which indicate that our selection does not introduce a noticeable bias in terms of the age distribution for clusters older than 10 Myr. The kinematical selection based on the RV uncertainty removed most of the very young clusters, the RV of young stars being less reliable.

\subsection{Kinematics of OCs compared with field stars}
In this section, we compare the kinematics of the field stars sample from \citet{kat18} with that of the OC sample. 

\begin{figure}[t]
\centering
\includegraphics[width=0.5\textwidth]{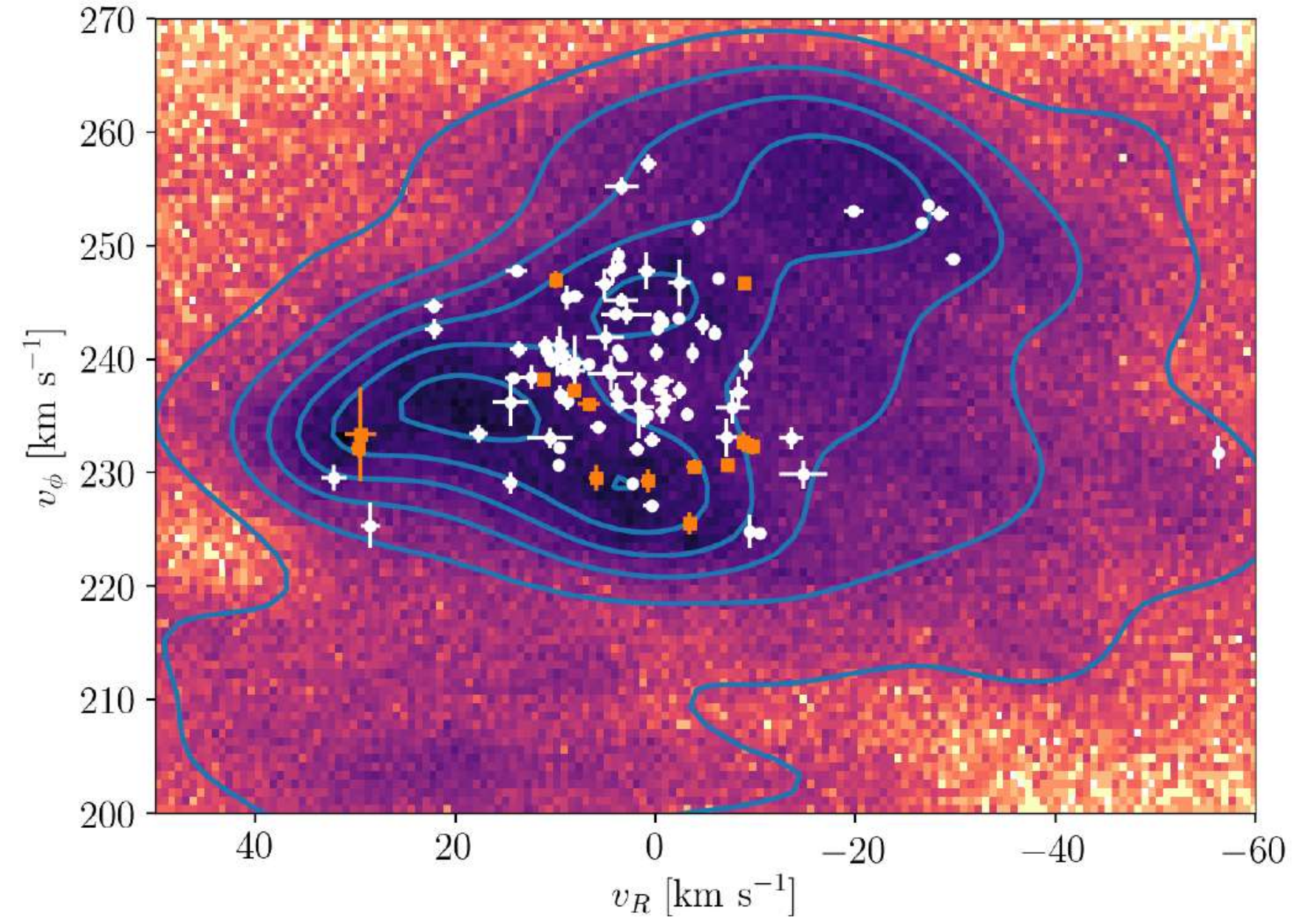}
\caption{($V_{\phi}$, $V_{r}$) distribution of field stars in the solar neighbourhood (closer than 200 pc) compared with clusters closer than 500 pc (white) and 200 pc (orange). 
}
\label{fig:vhpi_vr_field}
\end{figure}

In Fig.~\ref{fig:vhpi_vr_field} we compare the ($V_{\phi}$, $V_{r}$) distribution of the nearby OCs within 500 pc and 200 pc, with the field stars from \emph{Gaia} DR2 closer than 200 pc and having a relative error in parallax $<0.05$. In this space, field stars from the Solar neighbourhood show clear substructures in form of arches and clumps, which are associated with resonances due to non-axisymmetric features of the Galaxy, as discussed in \citet{kat18} and subsequent works. Large clumps were found to be associated with previously known moving groups \citep{egg58, deh98, che99, nor04}, and several overdensities of field stars in this space have been linked with OCs before \citep{kat18}. The general trend of our sample of clusters closer than 500 pc is to overlap with the higher density regions of field stars, as already seen in \citet{sou18}. The several OCs that we have gained here fall in the central part of the velocity distribution. Although we do not see a clear clumping of the clusters' distribution, it seems that OCs are most frequent in a band just between the two central arches of the field stars. Therefore, we see that OCs do not follow the exact distribution of overdensities drawn by the field stars. This was already seen with a smaller number of clusters in \citet{kat18}, where they mention that only the Pleiades and Hyades clusters are associated with overdensities, and other clusters do no show a particular over-density in this space.

\begin{figure}[t]
\centering
\includegraphics[width=0.5\textwidth]{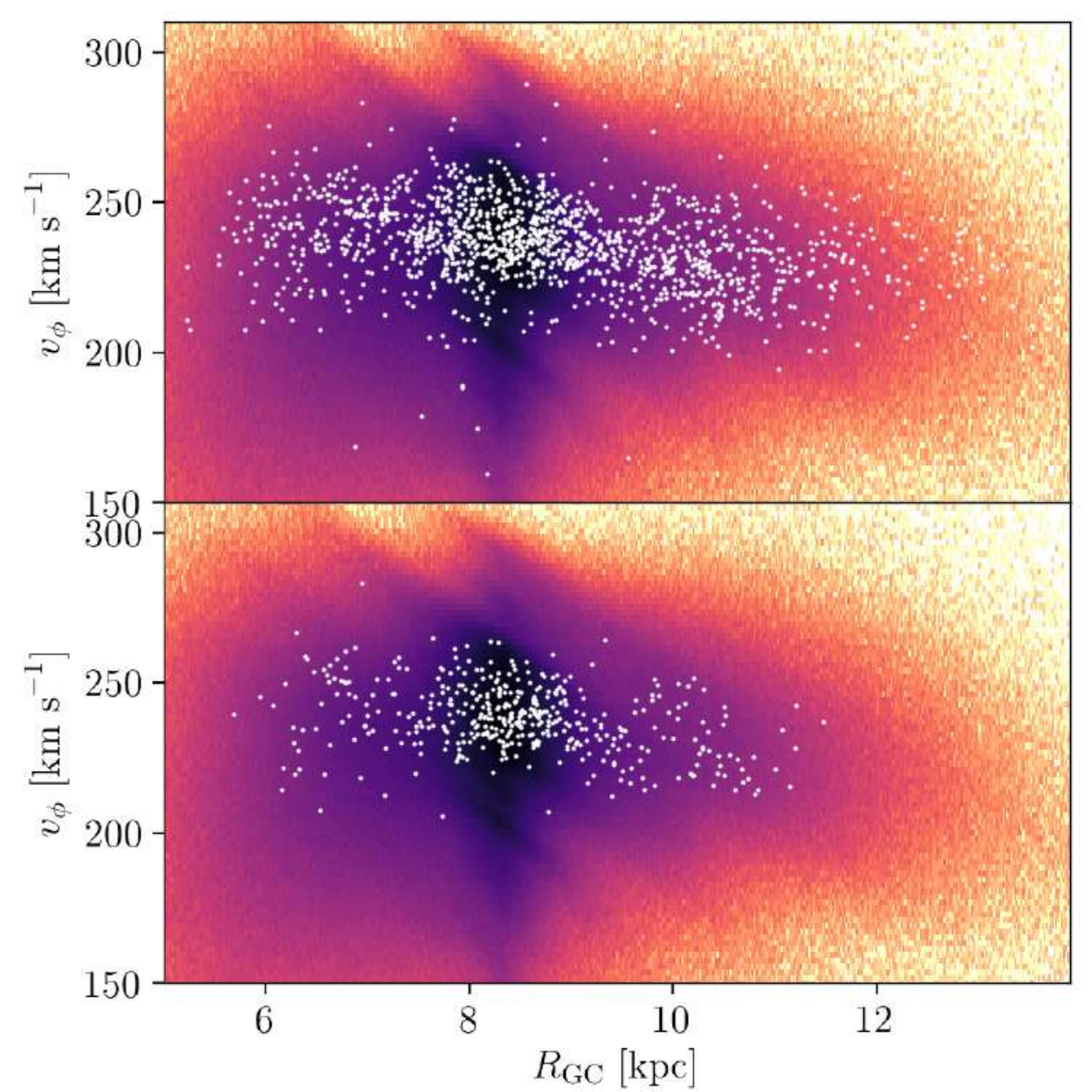}
\caption{($V_{\phi}$, $R_{\mathrm{GC}}$) distribution of field stars coloured by density, compared with clusters in white. In the top panel we plot the full sample of clusters, and in the bottom panel the HQS.}
\label{fig:vhpi_R_field}
\end{figure}

In Fig.~\ref{fig:vhpi_R_field} we plot the distribution of azimuthal velocities as a function of Galactocentric radius of the full sample of OCs and the HQS. Field stars in this figure were retrieved from the \emph{Gaia} archive selecting stars with the same criteria as \cite{kat18}, providing a sample of 6,376,803 stars.
In this space, \citet{ant18} showed that the diagonal ridges can be signatures of phase-mixing after a perturbing event, or alternatively due to resonances of a barred potential or of the spiral arms. \cite{kaw18} argued that many of the diagonal ridges are likely related with the perturbations from the bar outer Lindblad resonance and spiral arms, which may be explained with the transient spiral arm scenario. \cite{dia19} and \cite{bar20} claimed spiral resonances are able to trap stars or OCs orbits inside the corotation radius and associated the corotation to some ridges. In our plot, the clusters seem to qualitatively follow the distribution of diagonal overdensities described by the ridges, at least the most prominent ones. The recent N-body simulations by \citet{kha19} suggest that the ridges are more prominent for stars close to the plane with solar metallicity, conditions generally accomplished by a sample of OCs. Probably due to the shape of these ridges, we see a decreasing velocity gradient of $V_{\phi}$ towards large radii, more remarkable in the plot where all the sample of clusters is shown. This is consistent with what is seen in the top middle panel of Fig.~\ref{fig:velocities_age}, where outer clusters (yellow) have smaller velocities (and in general have older ages, see the histogram). This fact, as discussed more in detail in the next subsection, produces a dip in the rotation curve traced by our sample of OCs.

\subsection{Rotation curve of the Milky Way}\label{sec:rotcurve}
\begin{figure*}[h]
\centering
\includegraphics[width=\textwidth]{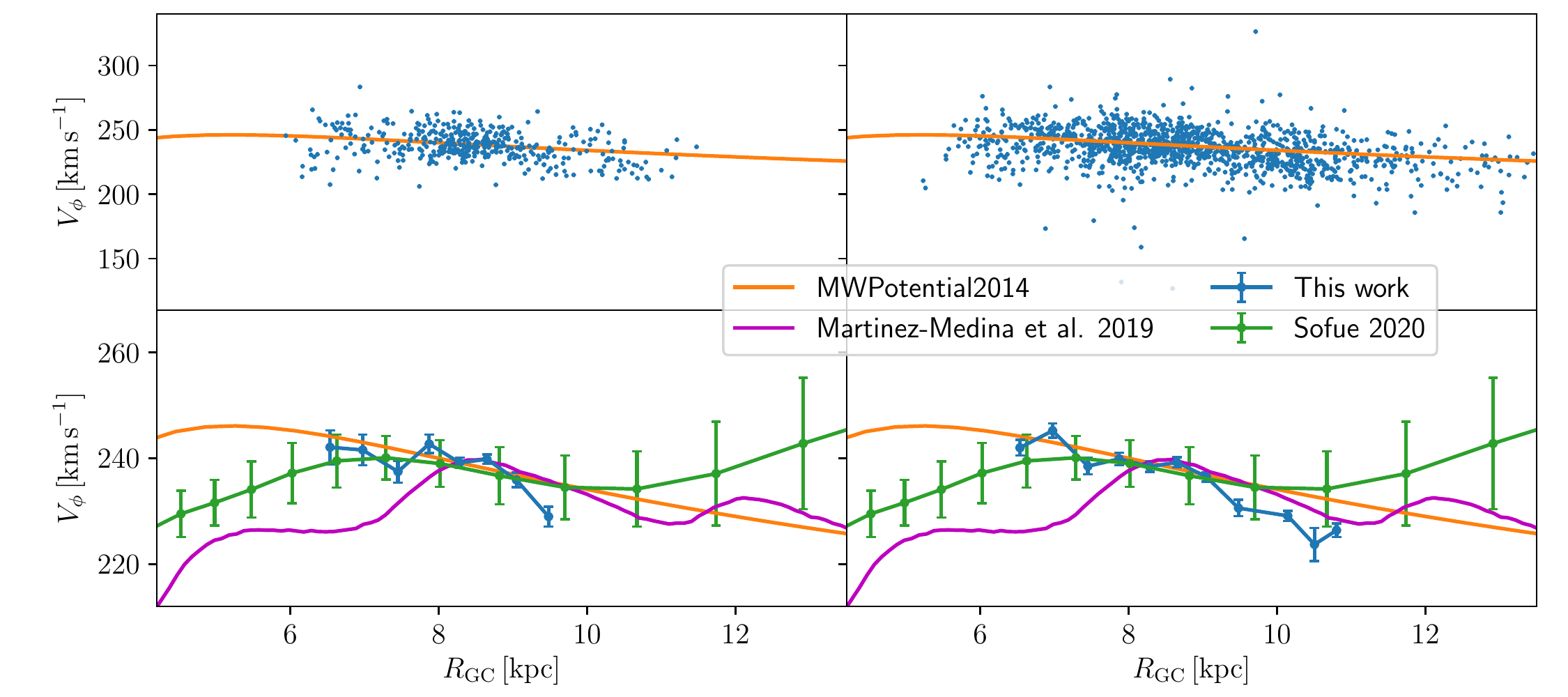}
\caption{Top: Azimuthal velocities of all the OCs in the full sample (right) and the HQS (left) as a function of $\rgc$. The rotation curve of the gravitational potential in Sect. \ref{action_angles_potentials} is superimposed. Bottom: median azimuthal velocities in bins of $\rgc$. We also superimpose the unified rotation curve computed by \citet{sof2020}, and the non-axisymmetric model by \citet{mar2019}.}
\label{fig:rotCurve}
\end{figure*}

We represent in Fig. \ref{fig:rotCurve} the azimuthal velocities of the full sample and the HQS OCs as a function of the Galactocentric radius, superimposing a theoretical rotation curve extracted from the axisymmetric gravitational potential described in Sect. \ref{action_angles_potentials}. Our sample of OC data describes with good precision the rotation curve of the Milky Way for a mostly young population in the range $\rgc\sim$[6.5,10.5] kpc.
In the bottom panels, we represent the median value of the rotation velocity in bins of 400 pc, approximately. Uncertainties are computed as $\sqrt{\nicefrac{\pi}{2}}\cdot\nicefrac{\sigma}{\sqrt{N}}$. We only plot the points computed using more than 20 OCs, which we consider reliable estimations and not dominated by outliers.

Both the HQS and the full sample show similar behaviour. We see a general overlap with what is expected from the axisymmetric curve ("MWPotential2014"). Several bins in both samples depart from the theoretical curve showing a small dip towards the inner Galaxy ($\rgc\sim$7 kpc), and a significant dip at $\rgc\sim$9.7 kpc, which departs more than $2\sigma$ from the theoretical curve. In the full sample, the outer dip extends up to 11 kpc. In the study by \citet{kat18}, similar dips seem to be present in the sample of small Z (their fig. 13), but the inner one is only seen at positive azimuths.
In the bottom panels, we overplot the unified rotation curve computed by \citet{sof2020}, which was obtained from a combination of data from different sources in the literature. Interestingly, this data shows a significant drop around 10 kpc, although with a different depth compared to our values.

There are several previous mentions in the literature to a dip in the region $\rgc\sim9.5$ kpc \citep{rei2019,bar2016,sof2009} whose nature is not fully understood. It has been tentatively explained by a ring density structure observed in the neutral H gas a bit further than the solar radius \citep[$\rgc\sim$10 kpc][]{nak2016}, which successfully reproduces the dip. \cite{dia19} calculated a corotation radius of $8.51 \pm 0.64$ kpc and found a similar dip in their rotation curve which can therefore be attributed to the effect of corotation, clusters located outside the corotation radius having a slower azimuthal velocity than clusters located inside. The depth of the dip is different depending on the data used, the one we observe here is compatible with the recent study by \citet{rei2019} from star-forming regions: a small decrease from the model of $\sim5$ \kms. Moreover, the general decreasing slope in the outer Galaxy shown by our data is also seen with the star-forming regions. Other studies from Cepheids \citep[e.g.][]{mro2019} also show a slight decrease in $V_{\phi}$ before and after the solar radius. \citet{mcg2019} built a model that reconciles the observed stellar rotation curve with the one seen by interstellar gas, taking into account the overdensities of the spiral arms. In their Fig. 4, the modelled points (green) show a clear dip inside the solar radius, compatible with the one we see in the cluster data in Fig.~\ref{fig:rotCurve}.

Comparing with Fig. \ref{fig:vhpi_R_field}, these dips are a direct consequence of the shape of the most prominent diagonal ridges, clearly shown by field stars in the \gaia\ RVS, but also by our clusters, where a remarkable decreasing velocity gradient is seen towards the outer Galaxy. This effect is analysed in \citet{mar2019} using a non-axisymmetric Galactic model, that shows that the duo bar-spiral arms produce diagonal-like ridges of stars of constant angular momentum. They show how these structures tend to clump the stars in the $V_{\phi}-\rgc$ plane, pulling them up at the beginning of the ridge, and down at the end. This translates into wiggles in the rotation curve of the Galaxy. We overplot in Fig. \ref{fig:rotCurve} the rotation curve derived from the model by \citet{mar2019}\footnote{we have scaled the rotation curve of the model to fit the reference frame used in our work, described in Sect.~\ref{sec:GalacticVelocities}, using the expression: $V_{\rm new} = V_{\rm old} + \frac{R_{\odot}}{8.5}(\theta_{\odot} - 214)$, also used in \citet{mro2019}}. Interestingly, the model resembles the two dips we observe with the OC data. We attribute to the same phenomenon the dips we observe in the cluster data, thus explained by the effect of non-axisymmetric structures in the angular momentum of the stars.



\subsection{Age dependence of Galactic velocities}\label{subsec:galactic_velocities}

\begin{figure}[t]
\centering
\includegraphics[width=\textwidth/2]{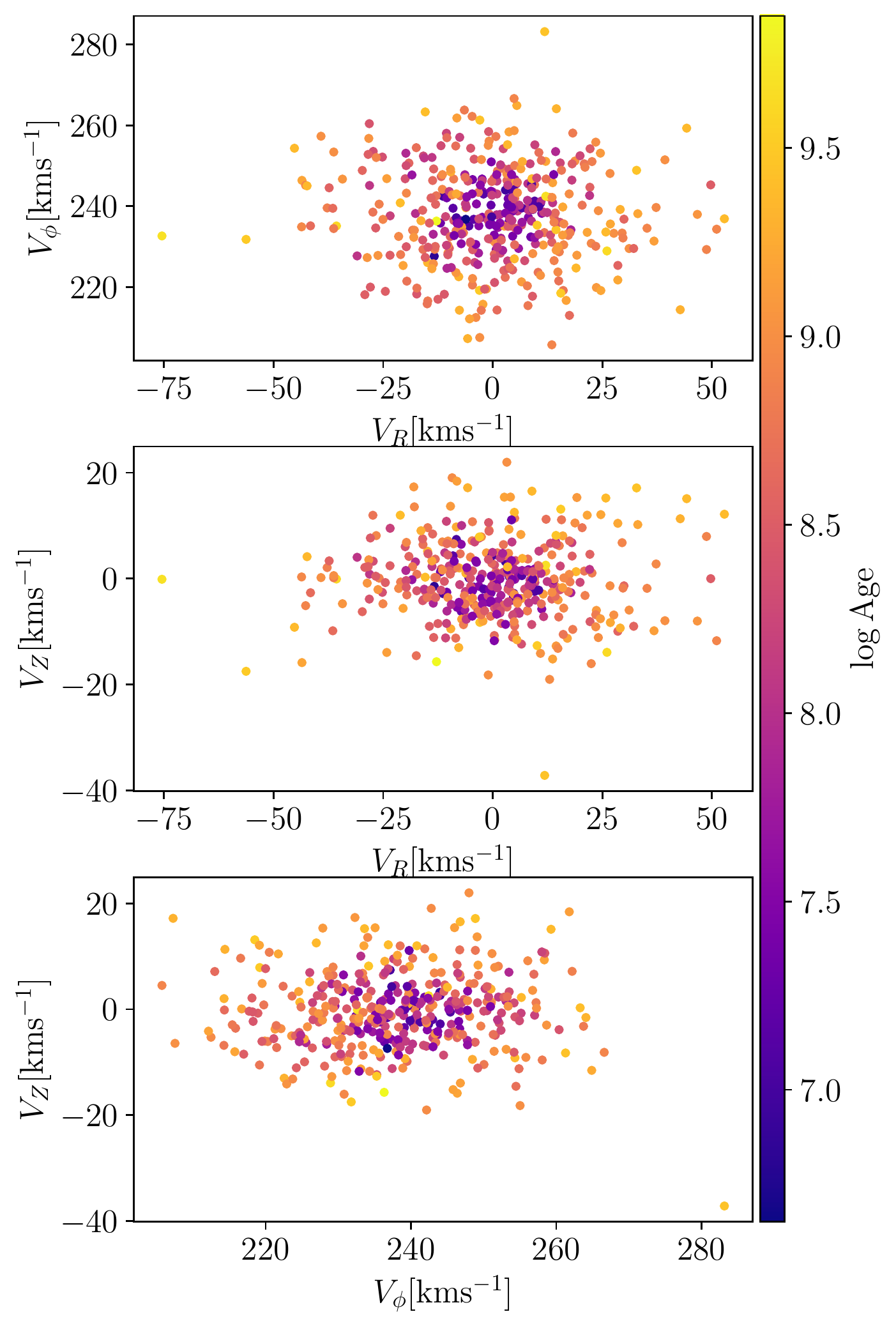}
\caption{Top: distribution of the HQS in Galactic velocities ($V_{r},V_{\phi}$) with colour related to age. Middle : same with the velocities ($V_{r},V_{Z}$). Bottom : same with the velocities ($V_{\phi},V_{Z}$).}
\label{f:distrib_velocities}
\end{figure}

Figure \ref{f:distrib_velocities} shows the distribution of the HQS in the ($V_{r},V_{\phi}$), ($V_{r},V_{Z}$) and ($V_{\phi},V_{Z}$) planes coloured by age. Several distant old OCs with extreme velocities ($V_{r} > 40$ \kms) have been recently discovered. Those included in the HQS are LP 930, UBC 326 and UBC 324 all located further than 1 kpc in heliocentric distance. The most extreme OCs in total velocity ($\sqrt{V_{r}^{2}+V_{\phi}^{2}+V_{z}^{2}}$) are Ruprecht 171, Haffner 5 and Berkeley 32 already known as high velocity objects for the disc.

 
\begin{figure*}[t]
\centering
\includegraphics[width=\textwidth]{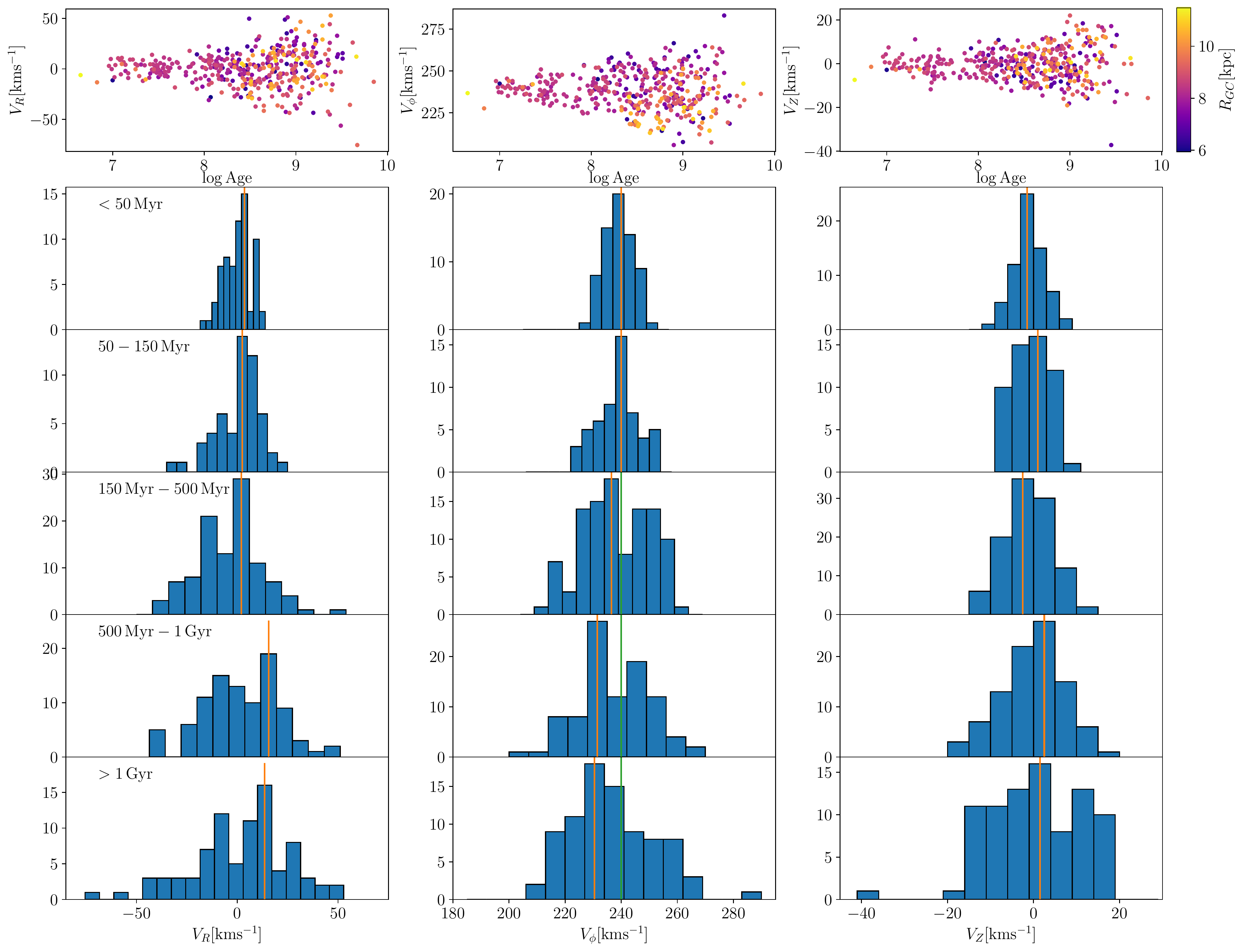}
\caption{Top: Radial, azimuthal and vertical velocities $V_{r},V_{\phi},V_{Z}$ as a function of the age, for the HQS with the Galactocentric radius shown with colours as indicated in the colour bar. Bottom: the five panels show the distribution of $V_{r},V_{\phi}$ and $V_{Z}$ of the OCs in different age bins. In each panel, the mode of the distribution is marked with an orange vertical line while the azimuthal velocity of the LSR is marked with a green vertical line in the middle column.}
\label{fig:velocities_age}
\end{figure*}

The three components of the velocities of the HQS clusters as a function of age are shown in the top panel of Fig. \ref{fig:velocities_age}. Clearly, they show a significant growing spread as age increases, which is related with the kinematic heating and will be discussed in more details in Sect \ref{subsec:avr}. The radial and vertical velocities are centred on zero while the azimuthal velocity is centred on the azimuthal velocity of the LSR. In addition, we can see in the middle panel of the top row that the azimuthal velocity of the oldest clusters within a given age range decreases when the radius increases. This trend is only noticeable for old clusters. Indeed, due to our kinematical selection based on the RV quality and errors of the Galactic velocities, most of the young distant clusters of our sample were removed. This bias was previously highlighted by \citet{kat18}: RVs for young stars are often less reliable due to their high temperature, their rotation or due to the fact that they are closer to the Galactic plane and not observable in spectroscopy due to heavy extinction. We can also see that inner clusters (in blue) are preferentially old, with a wide range of azimuthal velocities, a dozen of them having $V_{\phi} < 220$ \kms. For the vertical velocity shown on the right panel of the top row, the increase of the dispersion highlights the fact that the clusters reaching higher $V_{Z}$ are in general the oldest ones.

In the bottom panels of Fig. \ref{fig:velocities_age} is shown the histograms of each component of the velocities in several bins of age. The youngest clusters form a quite symmetric distribution, centered near zero for the radial and vertical components (median value of $V_{r} = -0.3$ \kms and $V_{Z} = -1.3$ \kms) and near the azimuthal velocity of the LSR for the azimuthal component (median value of $V_{\phi} = 239.5$ \kms). In the radial component, as age increases, the distribution becomes less gaussian and the peak of the histogram tends to move towards larger $V_{r}$, although this is dependent on the binning of the histogram and the taken age limits. Also depending on the binning, there seems to be a sign of bimodality in the 4 older age bins, although low number statistics does not allow to reach a significant conclusion. However, independently of the binning, the oldest bin has a significantly higher maximum value (around $\sim14$ \kms) with respect to the two youngest ones. Also, as age increases, the distribution becomes more and more asymmetric, with a large tail towards negative $V_{r}$, and a smaller tail on the positive side becoming more populated at older ages. This effect is also independent of the histogram binning and age limits. This implies that clusters moving inwards do it with a larger range of radial motions than those moving outwards. This trend is noticeable in the whole range of ages represented in our OC population, which is representative of the youngest population in the Galactic disk. The majority of old clusters of our sample have a positive $V_r$ highlighting that OCs which are moving outwards have more chance of survival than clusters moving inwards. For the azimuthal component, the histograms in bins of ages clearly show the asymmetric drift: in the two youngest bins the distribution is centred nearly on 240 \kms\ while the mode of the histogram shifts toward lower values in the older bins. In the three oldest bin, the most probable value is around 230 \kms. In the third and fourth age bins, we can also note a bimodality of the distribution also seen for the radial component of the velocity, the bimodality being however independent of the binning here. As age increases, the growing spread of the vertical velocity is clearly visible on the histograms, particularly in the two last age bins, where the distribution becomes asymmetric but remains centred near zero.



\begin{figure*}[h]
\centering
\includegraphics[width=\textwidth]{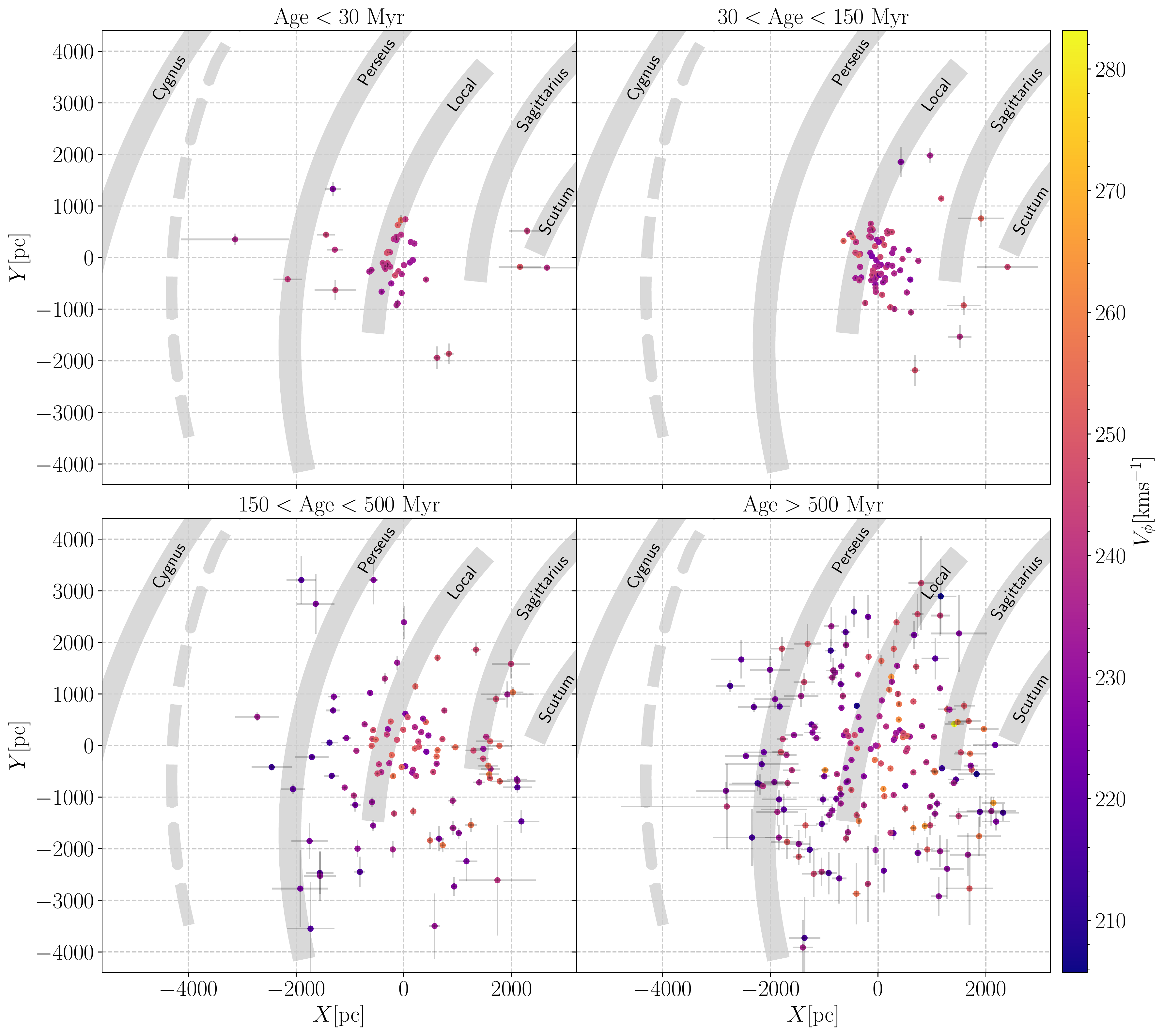}
\caption{Distribution of the HQS OCs in heliocentric cartesian coordinates. Colours stand for the azimuthal velocity $V_{\phi}$. The shaded structures show the spiral arm modelled by \citet{rei2014} while the dashed shaded structure represents the updated Cygnus arm from \citet{rei2019}.}
\label{fig:xy_vphi}
\end{figure*}

Figure \ref{fig:xy_vphi} shows the distribution of $V_{\phi}$ of the HQS sample in the X-Y plane in four different age bins. The spiral arms modelled by \citet{rei2014} are represented by the shaded structures while the updated Cygnus arm from \citet{rei2019} is represented by the dashed shaded structure. We note that in the two youngest bins, OCs are mainly located close to the Sun near the local arm and this spatial distribution is very different from that of the original sample presented in \cite{can20b} (their Fig. 8). This is a consequence of the bias of our sample stated above. The remaining nearby young clusters show a remarkable homogeneity in azimuthal velocities, with a typical velocity dispersion of ~5.5 \kms\ around the Sun velocity. In the two oldest age bins, OCs are much more spatially and kinematically dispersed and do not follow the spiral arms any more.


\subsection{Age velocity relation for open clusters}\label{subsec:avr}

The increasing trend of the velocity dispersions for field stars with age is known as the age-velocity relation (AVR) and has been discussed for decades \citep[e.g.][]{wie77,fre87,nor04}. The AVR is reasonably described as a power law with some saturation at the oldest ages \citep{bin08, aum09, sou08}, altough there is a debate about the shape of the relation \citep{mar14}. The AVR involves different mechanisms, such as the radial migration and heating induced by giant molecular clouds, the spiral arms \citep{mac19} or by the bar \citep{gra2016}, which affect differently the velocity components. The AVR study is complicated by the mixture of different stellar populations present in the solar neighbourhood, such as the thin and thick discs, which have different scale heights and different star formation and heating histories \citep{yu18,mac19}. Similarly, inner and outer populations are differently affected by the dynamical effect of the bar, the spiral arms and accretion events.  In addition,  uncertainties in the determination of stellar ages strongly affect the AVR \citep{aum16}. 
Ages of OCs are more reliable than ages of individual stars and determined homogeneously in \cite{can20b}. This allowed us to investigate the AVR of OCs in a large age range up to $\sim$2.5 Gyr (too few clusters older than this limit) and with a significant number of objects below 1 Gyr. The spatial extension is mostly confined within 200 pc from the Galactic plane, so the OC population is representative of the thin disc. It is thus interesting to compare the velocity ellipsoids of OCs and field stars in the age range where they overlap.

\begin{figure}[t]
\centering
\includegraphics[width=\textwidth/2]{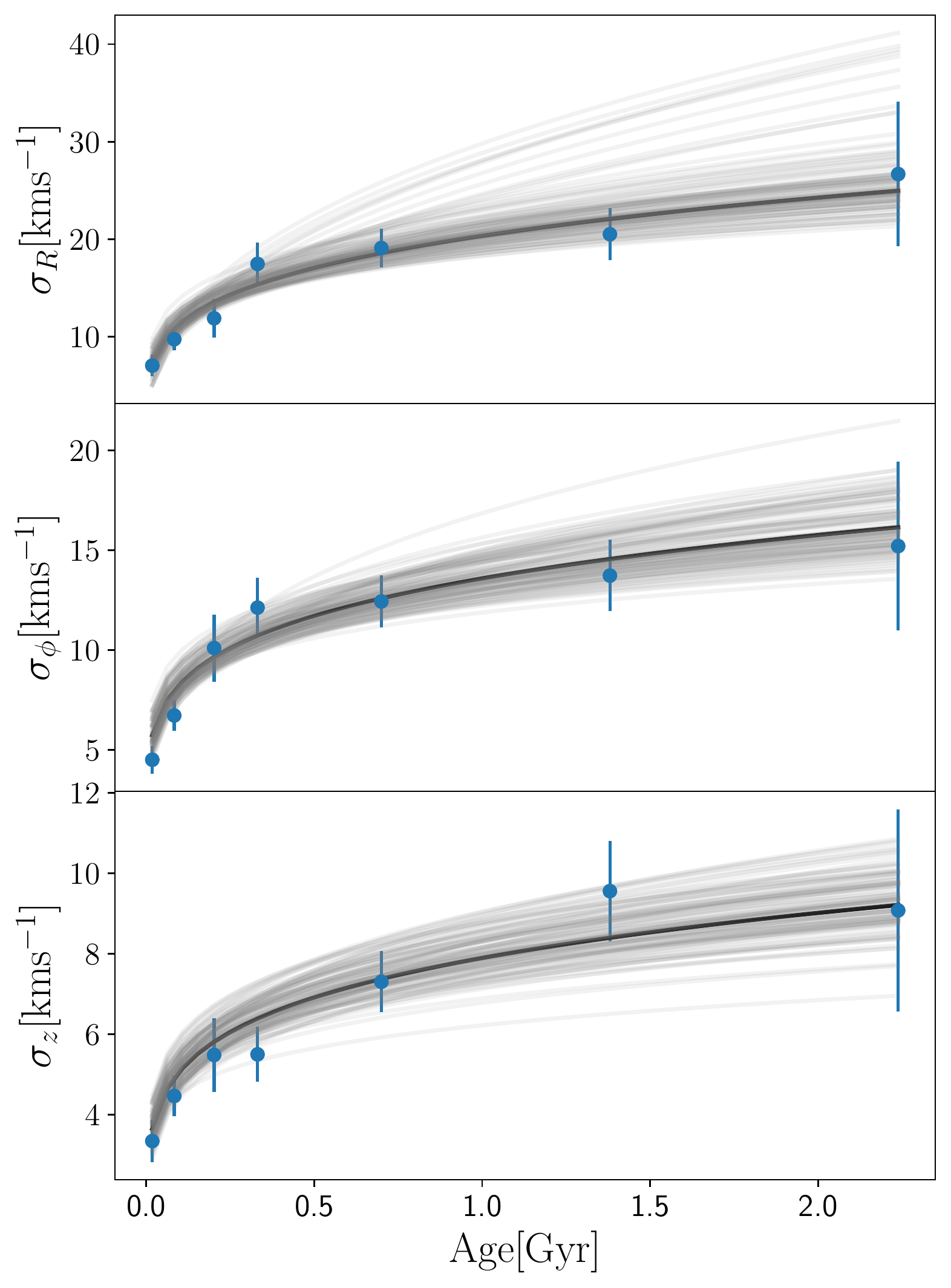}
\caption{Velocity dispersion of the cylindrical Galactocentric components $V_{r}$, $V_{\phi}$, $V_{z}$ for the HQS. The blue dots and the error bars respectively stand for the standard deviations of the velocities and their uncertainties obtained by fitting a gaussian on each component of the velocities in each of the 7 age bins of Table \ref{t:dispersions}, the oldest age bin having an upper limit of 2.5 Gyr. The x value of the blue dots is the median value of the corresponding age bin. The black line shows the best fit power law obtained by using a ML on each cluster of the HQS younger than 2.5 Gyr, and the grey lines represent the uncertainties of the fits.}

\label{fig:age_velocity_dispersion}
\end{figure}

Figure \ref{fig:age_velocity_dispersion} shows the velocity dispersions measured in seven age bins for the 418 OCs from the HQS with an age determination. 
In each age bins, we fitted a Gaussian to each component of the velocities. Each point represented in Fig. \ref{fig:age_velocity_dispersion} stands for these gaussian's standard deviations and their uncertainties. As expected, the dispersion of each component increases with age. The values and ratios are given in Table~\ref{t:dispersions}. The OC kinematics is characterised by a clear anisotropy in the 3 components, at all ages. The radial dispersion $\sigma_{R}$ is always significantly larger than $\sigma_{\phi}$ which is significantly larger than $\sigma_z$, as also seen in local field stars \citep[e.g.][]{bin08,kui94,ang18}. The dispersion ratios remain globally stable in the different age bins, with some fluctuations within the error bars. This implies that the velocity ellipsoid keeps the same shape at all ages. Following previous studies of field stars, we fitted the increase of the velocity dispersion with age ($\tau$) in the form of a power-law $\sigma_V \propto \tau^\beta$ as suggested in Sect. 8.4 of \citet{bin08} and by \citet{jen1992}. We used a Maximum Likelihood (ML) estimator on the individual clusters of the HQS (ages younger than 2.5 Gyr), assuming Gaussian velocity errors. The results of the fits are shown in Fig \ref{fig:age_velocity_dispersion}. The grey lines represent the uncertainties on the fits: we show 100 fits taken from the final sample of the ML. The black solid line shows the best fit power law obtained with a ML, this best fit is defined as the median value of the 16000 fits performed. 
We found $\beta_R=0.25^{+0.05}_{-0.03}$, $\beta_\phi=0.23^{+0.03}_{-0.03}$, $\beta_z= 0.19^{+0.03}_{-0.03}$. These values show that the heating rate of OCs is about the same in all directions, compatible with the velocity ellipsoid keeping the same shape at all ages as pointed earlier. In the following, we compare this kinematical behaviour with that of field stars.


 \begin{table*}[h]   
  \centering 
  \caption{Dispersions of velocity components and ratios in the same bins as in Fig. \ref{fig:age_velocity_dispersion}}
  \label{t:dispersions}
\begin{tabular}{lrrrrrrr}
\hline
 age interval (Gyr) & N & $\sigma_{R}$ & $\sigma_{\phi}$ & $\sigma_{z}$ & $\sigma_{R}/\sigma_{\phi}$ & $\sigma_{R}/\sigma_z$ & $\sigma_{\phi}/\sigma_{z}$ \\ 
\hline
age$<0.03$ & 43 & 7.05 $\pm$ 1.10 & 4.52 $\pm$ 0.71 & 3.34 $\pm$ 0.52 & 1.56 $\pm$ 0.49 & 2.11 $\pm$ 0.66 & 1.35 $\pm$ 0.42 \\
$0.03-0.15$ & 79 & 9.75 $\pm$ 1.11 & 6.73 $\pm$ 0.77 & 4.46 $\pm$ 0.51 & 1.45 $\pm$ 0.33 & 2.19 $\pm$ 0.50 & 1.51 $\pm$ 0.34 \\
$0.15-0.25$ & 38 & 11.90 $\pm$ 1.98 & 10.11 $\pm$ 1.68 & 5.48 $\pm$ 0.91 & 1.18 $\pm$ 0.39 & 2.17 $\pm$ 0.72 & 1.84 $\pm$ 0.61 \\
$0.25-0.50$ & 67 & 17.48 $\pm$ 2.17 & 12.13 $\pm$ 1.50 & 5.50 $\pm$ 0.68 & 1.44 $\pm$ 0.36 & 3.18 $\pm$ 0.79 & 2.21 $\pm$ 0.55 \\
$0.50-1.00$ & 94 & 19.11 $\pm$ 1.99 & 12.44 $\pm$ 1.30 & 7.30 $\pm$ 0.76 & 1.54 $\pm$ 0.32 & 2.62 $\pm$ 0.55 & 1.70 $\pm$ 0.36 \\
$1.00-2.00$ & 61 & 20.51 $\pm$ 2.67 & 13.73 $\pm$ 1.79 & 9.55 $\pm$ 1.24 & 1.49 $\pm$ 0.39 & 2.15 $\pm$ 0.56 & 1.44 $\pm$ 0.37 \\
$2.00-2.50$ & 15 & 26.68 $\pm$ 7.40 & 15.21 $\pm$ 4.22 & 9.07 $\pm$ 2.52 & 1.75 $\pm$ 0.97 & 2.94 $\pm$ 1.63 & 1.68 $\pm$ 0.93 \\
\hline
\end{tabular}
\end{table*}

\begin{table*}[h]
  \centering 
  \caption{Comparison between the values of $\beta$ found here and in previous studies}
  \label{t:beta_index}
\begin{tabular}{lrrr}
\hline
  & $\beta_{R}$ & $\beta_{\phi}$ & $\beta_{z}$ \\ 
\hline
This study & $0.25^{+0.05}_{-0.03}$ & $0.23^{+0.03}_{-0.03}$ & $0.19^{+0.03}_{-0.03}$ \\
\citet{yu18} & $0.28 \pm 0.08$ & $0.30 \pm 0.09$ &  $0.54 \pm 0.13$ \\
\cite{mac19} & - & - & 0.50 \\
\cite{sha20} & $0.251 \pm 0.006$ & - &  $0.441 \pm 0.007$ \\
\hline
\end{tabular}
\end{table*}




\cite{yu18} measured the AVR of $\sim$3\,500 local stars, splitting them into different subsamples depending on their Z position, and metallicity. Their metal-rich, low Z sample is comparable to our OC sample, representative of the local thin disc. For their two youngest age bins, corresponding to $1.4\pm0.4$ and $1.9\pm0.3$ Gyr respectively, they found dispersions in the different components and dispersion ratios in good agreement with what we found for OCs of similar age (see their Table 1). They fitted the heating parameters in the age range 1 to 8 Gyr and determined $\beta_R=0.28\pm 0.08$, $\beta_\phi=0.30 \pm 0.09$, $\beta_z= 0.54 \pm 0.13$, in agreement with our values in R and $\phi$, but not in Z.  

\cite{mac19} measured the heating parameters in R and Z with field stars of age between 1 and 9 Gyr. They found $\beta_z= 0.50$ for stars comparable to OCs in metallicity, higher than our value. They found $\beta_R$ to vary from 0.15  to 0.4 depending on the mean orbital radii. Recently \cite{sha20} assembled several large stellar samples tracing different populations from complementary large surveys and using different age estimators. They found consistent relations for all with $\beta_R=0.251\pm 0.006$, $\beta_z= 0.441 \pm 0.007$, so again in agreement with our value for R but not Z. \cite{mac19}  and \cite{sha20} highlighted a complex relation between $\sigma_R / \sigma_z$ with age,  depending on metallicity, angular momentum, and height above the plane.

These comparisons of the heating parameter $\beta$ (see Table \ref{t:beta_index} for more clarity) tend to show that the OC population has a dynamical evolution similar to the field stars in the radial and azimuthal directions, but not in the vertical direction. OCs seem to have a smaller heating rate in Z than field stars. The main difference between our determination of $\beta$ and that of \cite{yu18, mac19} and \cite{sha20} is the age range which extends to young ages for OCs (half of the OCs are younger than 360 million years) with few objects older than 3 Gyr, while the stellar samples typically range between 1 - 10 Gyr. Although their age distributions overlap around 1 - 2 Gyr, the fit of the $\beta$ parameter is not performed on the same age range for OCs and field stars. It is therefore interesting that the heating rate is found similar in the Galactic plane but not perpendicular to the plane. This could mean that clusters do not reach high altitude and older ages because they are disrupted before, introducing therefore a bias in our sample. The small $\beta_z$ could also reflect that giant molecular clouds which are the main cause of the vertical scattering of field stars \citep{lac1984, jen90} are not as efficient to scatter OCs, or that the effect of the giant molecular clouds is to disrupt the OCs. 

The heating of the Galactic disc and the destruction of clusters have been simulated among others by \cite{gus16}. They found that the fraction of massive old OCs, scattered into orbits with $|Z|>400$ pc, is typically 0.5\%. In the full initial sample of 2017 OCs from \cite{can20b}, 4\% have $|Z|>400$ pc. They are mainly old, with a median age of $\sim$2 Gyr, and at Galactocentric distances ranging from 7.9 to 20 kpc, as previously mentioned by \cite{can20b}. 
However, after our quality cuts, only 6 of such clusters remain in the HQS, making the statistics too poor to reach a conclusion on their origin.



\section{Actions and orbital parameters}\label{action_angles_potentials}
In this section we used the full 6D coordinates of the samples of clusters, to compute orbits and action variables, and analyze them as a function of the age.
We used the python package \texttt{galpy} \citep{galpy} for Galactic Dynamics to integrate the orbits and compute the action-angle variables.

We used the axisymmetric potential MWPotential2014 implemented in \texttt{galpy}, which was derived by \cite{galpy} fitting a simple model to existing dynamical data of the Milky Way. It is composed of a bulge, a Miyamoto-Nagai disc and a dark matter halo modelled by a NFW Potential. For the sake of comparison, we also used the axisymmetric potential model from \cite{mcm2017} to confirm our results. This model has also been fitted to the mass distribution of the Milky Way and is constituted of several components representing the cold gas disc close to the Galactic disc, both the thin and thick discs, a bulge and a dark matter halo. The detailed parameters of each component of these potentials are listed in \citet{galpy} and \citet{mcm2017}.

\subsection{Orbits}
\begin{figure}[h]
\centering
\includegraphics[width=\textwidth/2]{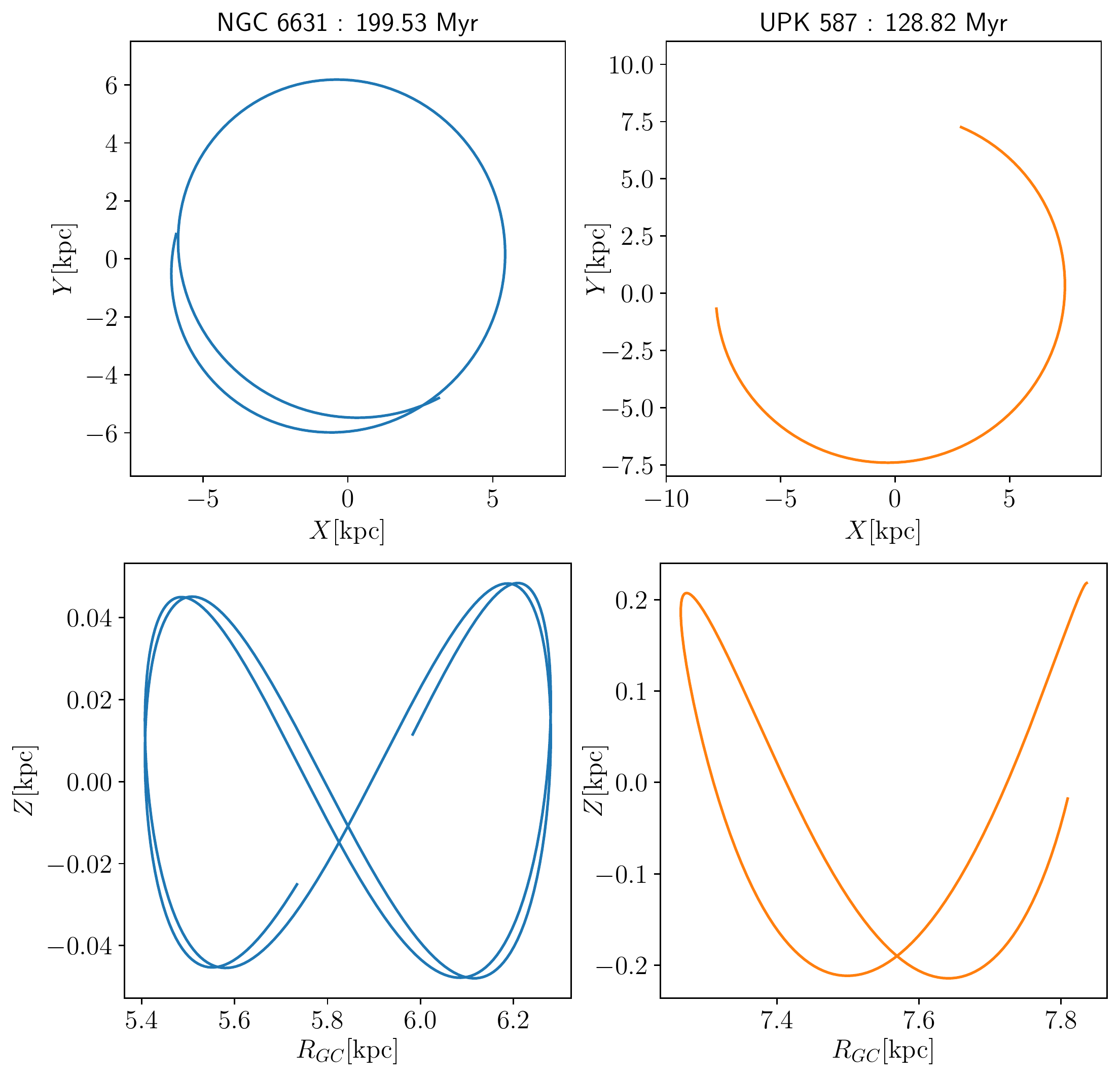}
\caption{Example of orbit for two OCs NGC 6631 and UPK 587 integrated with \texttt{galpy}.}
\label{fig:orbits}
\end{figure}

We integrated each cluster orbit with an integration step of 0.01 Myr years up to 500 Myr. We did not attempt to integrate more time (up to each cluster age) because the reliability of the results decreases significantly with time due to inaccuracies in the time-dependence of the potential, amplification of the uncertainties in distance and motions, among other effects \citep{hel2018}. We computed uncertainties integrating each orbit 1000 times with a Monte Carlo sampling in the same way as described in Sect.~\ref{sec:GalacticVelocities}. Figure \ref{fig:orbits} shows the orbits of two OCs, as an example of the results we can get using \texttt{galpy}.

\begin{figure*}[h]
\centering
\includegraphics[width=0.45\textwidth]{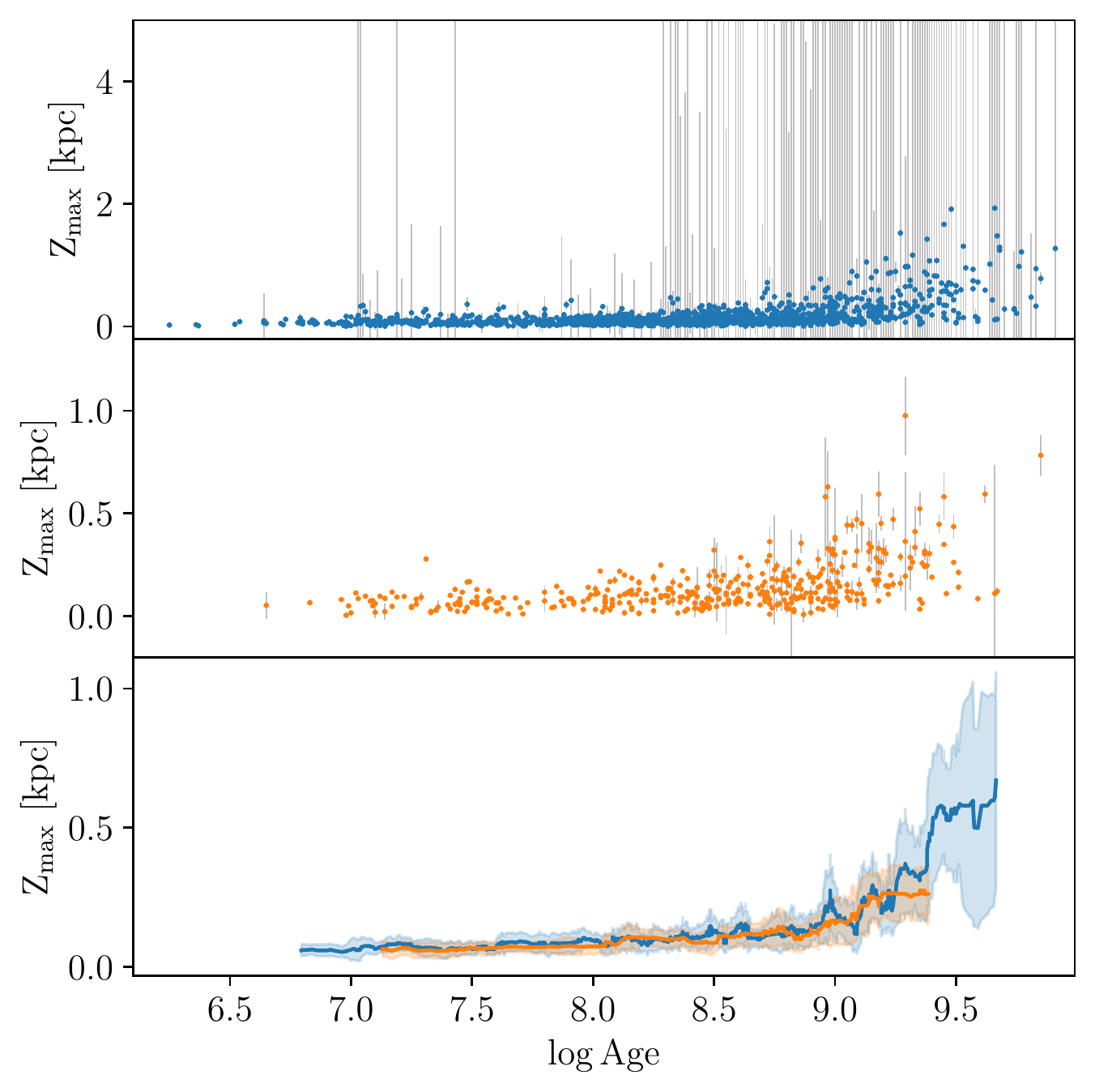}
\includegraphics[width=0.45\textwidth]{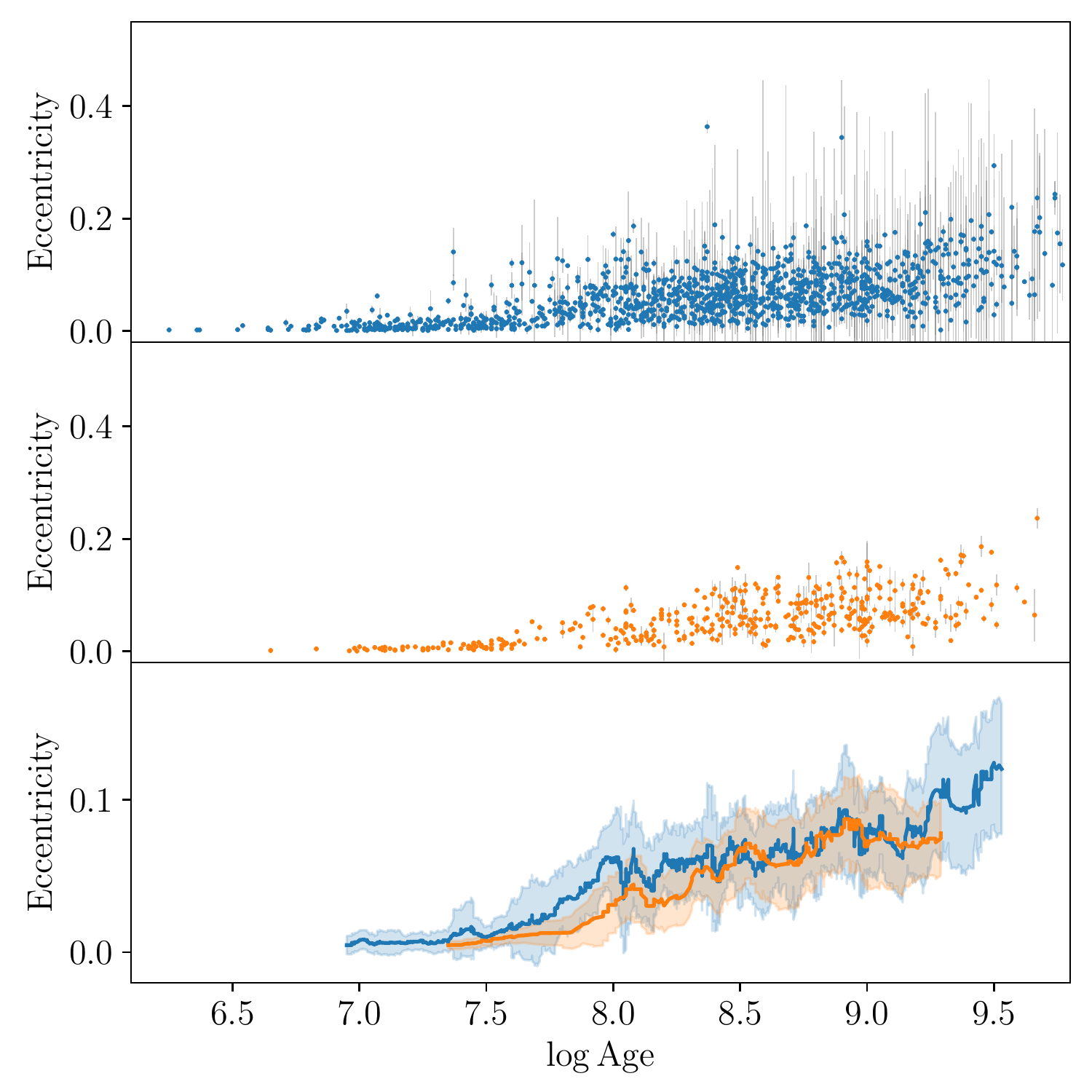}
\caption{Maximum altitude above the Galactic plane (left) and eccentricity (right) of the clusters for which we could integrate their orbits as a function of age. In both panels, we represented in blue (in the top panels) the full sample of OCs and in orange (in the middle panels) the HQS. The bottom panels shows the running median of the two samples, calculated with a window of 30 points. The shaded area corresponds to 1 MAD from the median (line). Note the change of scale between the top left panel and the two bottom left ones and between the two top right panels and the bottom right one.}
\label{fig:maximum_altitude_eccentricity}
\end{figure*}

We extract the orbital parameters of the 1315 OCs for which we have an age estimation to investigate their relation with age.
We represent in the left panel of Fig. \ref{fig:maximum_altitude_eccentricity} the evolution of the maximum altitude above the Galactic plane ($Z_{max}$) as a function of the age of the full sample of clusters and the HQS. In all the panels, the maximum height of clusters younger than 300 Myr remains constrained close to the Galactic plane. This is shown more clearly on the bottom panel where the running median of both samples increases only for an age higher than 1 Gyr. For both samples of OCs, the median age of the subsample of clusters which are reaching an altitude higher than 400 pc is greater than 1.5 Gyr. Also noticeable in the left panels of Fig. \ref{fig:maximum_altitude_eccentricity} is the increase of the dispersion of the maximum height of the OCs above the plane for ages older than 1 Gyr. This is usually attributed to the vertical heating of the disc: clusters are preferentially formed in the thin disc and then giant molecular clouds and spiral arms tend to scatter them away from the mid-plane \citep{spi1951, bin1990}. This effect is consistent with what is seen in more detail in the right panels of Fig.~\ref{fig:velocities_age}, already commented in Sect.~\ref{sec:GalacticVelocities}.

The eccentricity as a function of age in the right panel of Fig. \ref{fig:maximum_altitude_eccentricity} shows that : 
\begin{itemize}
    \item Clusters younger than 30 Myr ($\log (\mathrm{Age})\sim7.5$) show very low eccentricity. This is best seen in the HQS sample where the maximum value in this range is 0.018, while in the full sample there are some outliers with values up to 0.15. Very young clusters have therefore nearly circular orbits.
    \item For clusters older than 30 Myr the dispersion of the eccentricities at a given age is large for both samples. The running median of the HQS shows an increase which is quite smooth at least up to 300 Myr ($\log (\mathrm{Age})\sim8.5$). The full sample exhibits a similar behaviour to that of the HQS but with slightly larger mean values of eccentricities.
    \item For ages older than 1 Gyr the dispersion in eccentricities starts to be very large for the full sample. For the HQS there seems to be a stabilization of the eccentricities around a mean value of 0.08.
\end{itemize}

This shows that OCs are born on nearly circular orbits and as their age increases, they are more prone to suffer gravitational perturbations from non-axisymmetric components.
We do not find very young clusters with large eccentricities, but we find old clusters with both large and low eccentricities. We do not see a preferential location in the Galaxy, or differential characteristics between large and low eccentricity old clusters. 

\subsection{Action angle variables}

The action-angle variables are a set of canonical coordinates which are proved to be useful to study the substructure of stars in the 6D phase space. As extensively discussed by several authors \citep[e.g.][]{mcm08}, orbital actions $(J_R,J_{\phi},J_z)$ are integrals of motion in an axisymmetric potential, but they also provide information of non-axisymmetric perturbations. In addition, they are independent of time and are therefore more reliable to describe the orbital parameters by removing their time dependence. They have been used to describe stellar components in our Galaxy, in particular the moving groups seen in the solar neighbourhood in the $(U,V)$ plane \citep{tri19}.

In an axisymmetric potential, action variables can easily be interpreted as physical quantities. The radial action $J_{R}$ can be used as a proxy for orbit's eccentricity or as a measure of the oscillations around the guiding radius of the object. The azimuthal action $J_{\phi}$ is equal to the angular momentum in the vertical direction $L_{Z}$, which indicates the quantity of rotation of the object around the centre of the Galaxy. Similarly, the last coordinate, the vertical action $J_{z}$, can be used as a measure of the oscillations of the object around the plane of the Galaxy and therefore is a proxy for the maximum height of the object along its orbit.

All of these quantities are conserved in an axisymmetric potential but are affected by non-axisymmetric structures such as a bar, spiral arms or by a merger. We refer to Sect. 3 of \cite{bin08} for a mathematical and comprehensive description of these variables and their meanings.

We made the computation for the 411 clusters from the HQS with known age. For comparison purposes, we made the same computation with the sample of stars within $d<200$ pc in \emph{Gaia} DR2 RVS, i.e. the same selection as the one done by \citet{tri19}, which counts $\sim350,000$ stars.

\begin{figure*}[h]
\centering
\includegraphics[width=\textwidth]{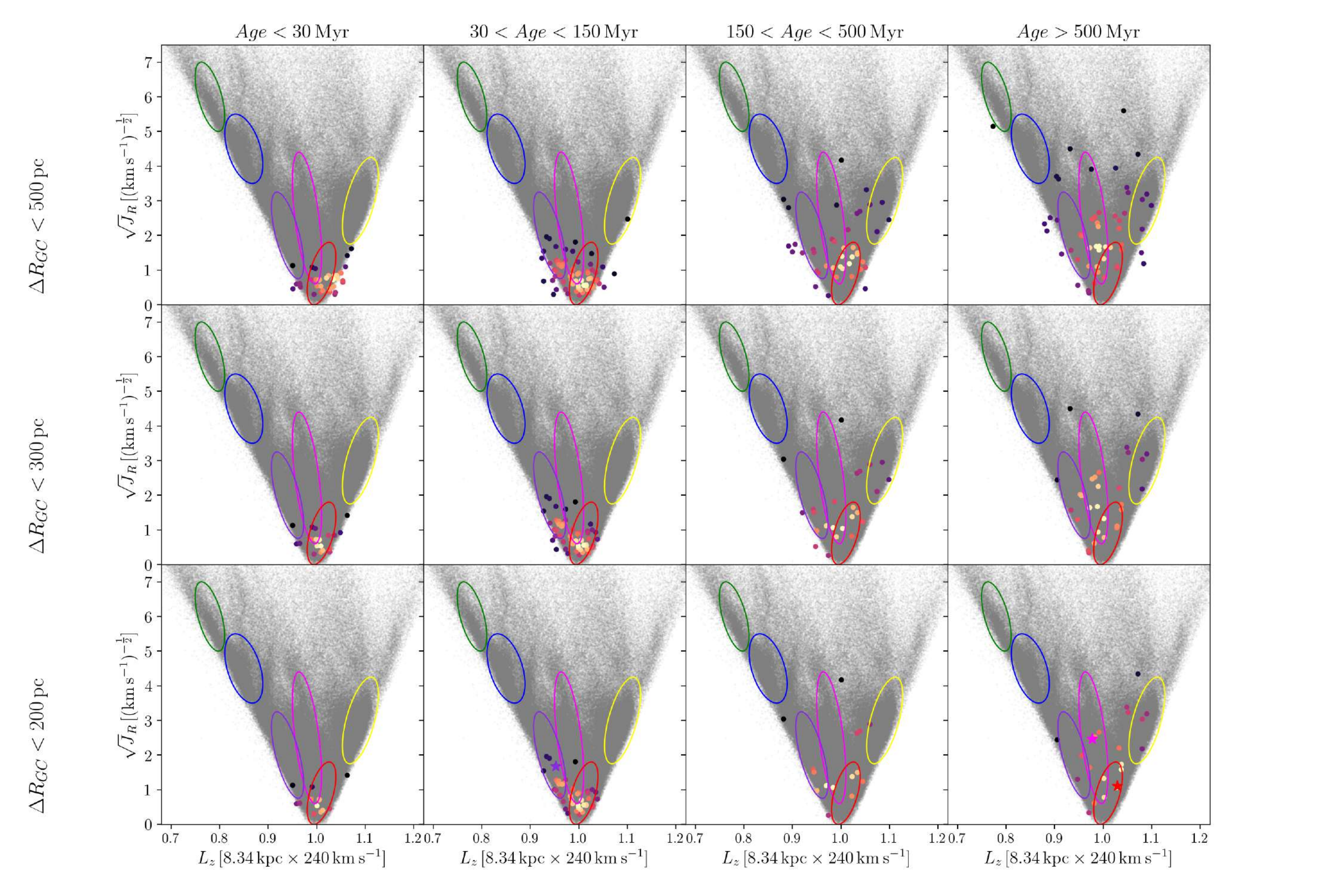}
\caption{Radial action vs angular momentum $(J_R,L_z)$ distribution of field stars closer than 200 pc in \emph{Gaia} DR2/RVS sample (grey), and clusters (coloured dots by the density of points). Each column shows clusters in different age bins, and each row includes clusters selected inside different ranges of Galactocentric radius, instead of heliocentric distances. We indicate the approximate location of the known moving groups analyzed by \citet{tri19}: Hercules (green and blue), Pleiades (dark violet), Hyades (magenta), Coma Berenices (red), Sirius (yellow). We highlight the values of the three clusters Pleiades, Hyades and Coma Berenices with a star of colour violet (second column of the bottom row), pink and red (both in the last column of the bottom row), respectively, in their age-$R_{\mathrm{GC}}$ panels.}
\label{fig:action_angle}
\end{figure*}

In Fig. \ref{fig:action_angle} we show the distribution of radial action $J_R$ with respect to the vertical component of the angular momentum $L_Z$, of field stars (in grey) compared with the sample of HQS clusters (coloured by the normalized density of points). In all subplots, the sample of field stars is the same as we don't know their ages, and we indicated with coloured ellipses the approximate location of the known moving groups present in the Solar neighbourhood \citep{ant08,kat18}, as analyzed in the action space by \citet{tri19}. In this space, the location of a star is interpreted in relation to its orbital characteristics as following: (i) the "V" shape is due to the cut in distance (200 pc) made for the sample selection, (ii) stars with circular orbits are placed at $(L_Z,\sqrt{J_R})\sim(1,0)$, while more eccentric orbits appear at larger $J_R$, (iii) stars close to their apocenter (pericenter) are placed in the left (right) edge of the "V".

We restrict the volume of analyzed clusters using their Galactocentric radius instead of their heliocentric distances, because of the reduced amount of clusters at $d<200$ pc. Even though this made the comparison between cluster and field more difficult, the precision in the distances and velocities of the clusters is much better than for individual field stars, so we expected that the kinematic information will not be blurred by this, as it happens usually for individual stars. We made the cuts in $R_{GC}=8.3 \pm [0.2,0.3,0.5]$ pc, showed in each row. Using the information of cluster ages we were able to add an additional dimension to the figure, so we dissected the sample in four age bins [$<30$,$30-150$,$150-500$,$>500$] Myr, showed as columns. The aforementioned differences in the volume selection of the clusters and the field stars are visible in the top row where some of the clusters stand out of the left/right edges, which would correspond to clusters towards the Galactic centre/anticentre.

From Fig.~\ref{fig:action_angle} we see that the bulk of clusters tends to be concentrated towards the position $(1,0)$, not surprising since most OCs have cold kinematics, and tend to have nearly circular orbits. 
It is visible in all three distance limits (rows) that there are few clusters at large eccentricities, but increasing in number at older ages. As a consequence, the density peak (marked in yellow) moves upwards towards older ages. This is particularly clear in the top row because of the large number of points.

Regarding the relation of the clusters and the moving groups, we see that there are no clusters populating the two Hercules substreams, which are stars with too eccentric orbits and low angular momentum. On the contrary, the location of the Hyades/Pleiades/Coma Berenices/Sirius moving groups resembles more the distribution of the clusters, even at larger distances (top row). We highlight that for the age bin $30-150$ Myr (for all distance cuts) there is a significant clump of clusters towards the mid-left side (smaller $L_z$), populating the region where the Pleiades moving group is found. On the other hand, the Sirius moving group is populated mainly by clusters in the two older age bins. These dependencies in the population in each moving group depending on age are in the same direction as the findings by \citet{ant08}: in their Fig. 14, Hyades-Pleiades-Coma gather most of their young stars ($\lesssim300$ Myr) and at $\gtrsim300$ Myr there is a bump of stars related to Sirius. Hercules is associated with stars of $\sim2$ Gyr, and the distributions of all moving groups also exhibit a bump at this age, however, we almost do not find clusters at this old age, so there is no indication of this in our sample.

\section{Summary}\label{sec:conclusion}
 Thanks to the combination of \gaia\ and ground-based surveys and catalogues, and with new memberships from \cite{can20b}, we assembled the largest catalogue of RV velocities for OCs in order to study their kinematics. As a by-product of our study, that includes the \emph{Gaia} DR2 RVS, \emph{Gaia}-ESO survey, APOGEE, RAVE, GALAH and smaller catalogues, we compared the RVs from the different sources to each other to assess their typical precision and zero-point. We found RV zero points to be consistent at a level better than 1 \kms. The scatter of the comparisons indicate that the real precision of each catalogue is compatible with the individual uncertainties listed in it.
All non-\gaia\ RV measurements were corrected to align them on the \gaia\ RVS zero point. The weighted mean RV of each star and each cluster resulted in 1\,382 OCs having a RV, 38\% with a highly reliable RV based on more than 3 stars and with an uncertainty lower than 3 \kms. 

We computed both heliocentric and Galactocentric cartesian and cylindrical velocities for this sample of OCs and defined a High Quality Sample composed of 418 OCs with the most reliable velocities out of which 411 OCs have an age determination. We found that most OCs fall in a band in between the two main arches drawn by field star in the $V_r-V_{\phi}$ plane, while they seem to follow the overdensities described by the diagonal ridges in the $R_{GC}-V_{\phi}$ plane. The rotation curve drawn by our OCs shows two significant dips: at $R_{GC} \sim 7$ kpc, and a more prominent one around $R_{GC} \sim 9.7$ kpc. The locations and depths of these dips are in agreement with the perturbations we would expect from the non-axisymmetric components of the disc which also draw the ridges observed in the $R_{GC}-V_{\phi}$ plane.

With the ages of almost all the clusters from our sample, we investigated in details the age velocity relation for OCs which shows a clear anisotropy between the 3 components of the velocities. Compared with field stars studies, the heating parameter $\beta$ of OCs was found to be similar in the radial and azimuthal directions, but significantly lower in the vertical direction. This low heating rate in the Z coordinate can be due to the disruption of old clusters which are the most likely to reach high altitudes above the disc or to a less efficient heating of OCs by giant molecular clouds. Although we are aware that the quality cuts we applied discarded distant clusters resulting in a bias of our sample. 

We use the 6D + age information of our sample of OCs to compute and investigate orbits and action variables. We analyze the dependencies of the recovered orbital parameters as a function of age. We see that most of the clusters reach a maximum altitude above the plane during their orbits smaller than 400 pc, and only those older than 1 Gyr are able to move considerably away from the midplane, but typically less than 1 kpc.  Clusters younger than 30 Myr show a very low eccentricity ($\sim0.018$), and for clusters older than that, especially those older than 100 Myr, the eccentricity shows an increasing relation with age. These results show that OCs are born in circular orbits, and as age increases, they are more prone to suffer perturbations of their orbits. This is also seen after the computation of action variables, where, as age increases, the distribution in the $(L_Z,\sqrt{J_R})$ plane tends to spread beyond $\sim(1,0)$. We relate our cluster distribution in this action space with the location of the known moving groups, as a function of age. We conclude that the Pleiades-Hyades-Coma moving groups seem to be more populated by young clusters, while the Sirius region seems to have a clump of clusters of age $\gtrsim300$ Myr. No clusters are populating the two Hercules streams.

\begin{acknowledgements}
This work has made use of data from the European Space Agency (ESA) mission \emph{Gaia} (\url{http://www.cosmos.esa.int/Gaia}), processed by the \emph{Gaia} Data Processing and Analysis Consortium (DPAC, \url{http://www.cosmos.esa.int/web/Gaia/dpac/consortium}). We acknowledge the \emph{Gaia} Project Scientist Support Team and the \emph{Gaia} DPAC. Funding for the DPAC has been provided by national institutions, in particular the institutions participating in the \emph{Gaia} Multilateral Agreement.
This research made extensive use of the SIMBAD database, and the VizieR catalogue access tool, operated at the CDS, Strasbourg, France, and of NASA Astrophysics Data System Bibliographic Services.
This research has made use of Astropy \citep{Astropy2013}, Topcat \citep{Taylor2005}.

Y.T., C.S., and L.C. acknowledge support from "programme national de physique stellaire" (PNPS) and from the "programme national cosmologie et galaxies" (PNCG) of CNRS/INSU. L.C. acknowledges the support of the postdoc fellowship from French Centre National d’Etudes Spatiales (CNES). This work has been supported by the Spanish Ministry of Economy (MINECO/FEDER, UE) through grants ESP2016-80079-C2-1-R, RTI2018-095076-B-C21 and the Institute of Cosmos Sciences University of Barcelona (ICCUB, Unidad de Excelencia ’Mar\'{\i}a de Maeztu’) through grant MDM-2014-0369 and CEX2019-000918-M. A.B. acknowledges support from Italian MIUR Premiale 2016 "MITiC". AM acknowledges the support from the Portuguese FCT Strategic Programme UID/FIS/00099/2019 for CENTRA. This project has received funding from the European Union’s Horizon 2020 research and innovation programme under the Marie Sk\l{}odowska-Curie grant agreement No. 800502. DB is supported by Fundação para a Ciência e Tecnologia (FCT) through national funds and by FEDER - Fundo Europeu de Desenvolvimento Regional through COMPETE2020 in connection to these grants: UID/FIS/04434/2019; PTDC/FIS-AST/30389/2017 \& POCI-01-0145-FEDER-030389. We warmly thank L. Mart\'inez-Medina for his comments and for sharing with us his models.

\end{acknowledgements}

%
%

\bibliographystyle{aa} 
\bibliography{biblio}
\begin{appendix}

\end{appendix}

\end{document}